# A GENERIC FINITE ELEMENT FRAMEWORK ON PARALLEL TREE-BASED ADAPTIVE MESHES

SANTIAGO BADIA[1,2], ALBERTO F. MARTÍN[1], ERIC NEIVA[2,3], AND FRANCESC VERDUGO[2]

ABSTRACT. In this work we formally derive and prove the correctness of the algorithms and data structures in a parallel, distributed-memory, *generic* finite element framework that supports *h*-adaptivity on computational domains represented as forest-of-trees. The framework is grounded on a rich representation of the adaptive mesh suitable for generic finite elements that is built on top of a low-level, light-weight forest-of-trees data structure handled by a specialized, highly parallel adaptive meshing engine, for which we have identified the requirements it must fulfill to be coupled into our framework. Atop this two-layered mesh representation, we build the rest of data structures required for the numerical integration and assembly of the discrete system of linear equations. We consider algorithms that are suitable for both subassembled and fully-assembled distributed data layouts of linear system matrices. The proposed framework has been implemented within the `FEMPAR` scientific software library, using `p4est` as a practical forest-of-octrees demonstrator. A strong scaling study of this implementation when applied to Poisson and Maxwell problems reveals remarkable scalability up to 32.2K CPU cores and 482.2M degrees of freedom. Besides, a comparative performance study of `FEMPAR` and the state-of-the-art `deal.II` finite element software shows at least comparative performance, and at most factor 2-3 improvements in the *h*-adaptive approximation of a Poisson problem with first- and second-order Lagrangian finite elements, respectively.

## 1. INTRODUCTION

Research over the past few years has led to a new generation of petascale-capable algorithms and software for fast Adaptive Mesh Refinement and coarsening (AMR) using adaptive tree-based meshes endowed with Space-Filling Curves (SFCs) [1]. SFCs are exploited for efficient data storage and traversal of the adaptive mesh, fast computation of hierarchy and neighborhood relations among the mesh cells, and scalable partitioning and dynamic load balancing. For hexahedral meshes, the state-of-the-art in tree-based AMR with SFCs is available at the `p4est` software [2, 3]. It provides parallel forest-of-octrees (i.e., a data structure which results from patching together multiple adaptive octrees) grounded on the standard 1:$2^d$ isotropic refinement rule and the Morton SFC index [2]. We refer to [4] (and references therein) for works on single-octree handling, and to [2] for their extension to a forest-of-octrees. However, tree-based AMR with SFCs can be generalized to other cell topologies as well. The authors of [5] present an approach for simplicial adaptive meshes that is grounded on Bey's red-refinement rule, and the so-called Tetrahedral Morton (TM) SFC index [5]. Holke's dissertation [1] goes even further by providing abstract, extensible forest-of-trees manipulation algorithms, i.e., cell-topology- and dimension-independent. By means of implementing a set of low-level local-to-cell operations, these algorithms may be extended to work with

[1]SCHOOL OF MATHEMATICAL SCIENCES, MONASH UNIVERSITY, CLAYTON, VICTORIA, 3800, AUSTRALIA.
[2]CIMNE – CENTRE INTERNACIONAL DE MÈTODES NUMÈRICS EN ENGINYERIA, UPC, ESTEVE TERRADAS 5, 08860 CASTELLDEFELS, SPAIN.
[3]DEPARTMENT OF CIVIL AND ENVIRONMENTAL ENGINEERING. UNIVERSITAT POLITÈCNICA DE CATALUNYA, JORDI GIRONA 1-3, EDIFICI C1, 08034 BARCELONA, SPAIN.
*E-mail addresses*: santiago.badia@monash.edu, alberto.martin@monash.edu, eneiva@cimne.upc.edu, fverdugo@cimne.upc.edu.

*Date*: April 3, 2020.

Financial support from the European Commission under the EMUSIC and ExaQUte projects within the Horizon 2020 Framework Programme is gratefully acknowledged (Grant Ids. 690725 and 800898, resp.). This work has been partially funded by the project MTM2014-60713-P from the "Ministerio de Economía, industria y Competitividad" of Spain. SB gratefully acknowledges the support received from the Catalan Government through the ICREA Acadèmia Research Program. EN gratefully acknowledges the support received from the Catalan Government through an FI fellowship (2019 FI-B2-00090; 2018 FI-B1-00095; 2017 FI-B-00219). FV gratefully acknowledges the support received from the *Secretaria d'Universitats i Recerca* of the Catalan Government in the framework of the Beatriu Pinós Program (Grant Id.: 2016 BP 00145). The authors thankfully acknowledge the computer resources at Marenostrum-IV and the technical support provided by the Barcelona Supercomputing Center (RES-ActivityID: FI-2018-2-0009, FI-2018-3-0029, IM-2020-1-0002).

general polytopes (e.g., lines, simplices, bricks, prisms, pyramids, etc.). These ideas are being implemented in the on-going, open source software effort `t8code` [1].

The exploitation of single-octree or, more generally, forest-of-octrees in parallel adaptive Finite Element (FE) solvers (see, e.g., [6]) has been shown to be cornerstone in several large-scale application problems (see, e.g., [7–9]). Forest-of-trees meshes provide multi-resolution capability by local adaptation. Indeed, this capability makes them perfectly suitable for the efficient approximation of multi-scale Partial Differential Equations (PDEs). They are also highly appealing for PDEs posed on complex geometries in combination with unfitted (a.k.a. embedded boundary) FE methods [10, 11]. However, multi-resolution comes at a price. Forest-of-trees meshes are, in the most general case, *non-conforming*, i.e., they contain the so-called *hanging Vertices, Edges and Faces (VEFs)*. These occur at the interface of neighboring cells with different refinement levels. Mesh non-conformity introduces additional implementation complexity specially in the case of conforming FE formulations. Degrees Of Freedom (DOFs) sitting on *hanging* VEFs cannot have an arbitrary value, as this would result in violating the trace continuity requirements for conformity across interfaces shared by a coarse cell and its finer cell neighbors.

The set up (during FE space construction) and application (during FE assembly) of hanging DOFs constraints is well-established knowledge; see e.g., [12] and [13], resp. In a parallel distributed-memory environment, these steps are much more involved. In order to scale FE simulations to large core counts, the adaptive mesh must be partitioned among the parallel tasks such that each of these only holds a fraction of the global mesh cells. In particular, an overlapping partition of the mesh cells is required where the processor-local portion is extended with a set of geometrically adjacent (neighboring) off-processor cells, i.e., the so-called *ghost* cells. The standard practice is to constraint the data structures to keep a single layer of ghost cells, thus minimizing the impact of these extra cells on memory scalability. Under this constraint, if a set of suitable conditions are not satisfied, then the hanging DOF constraint dependencies may expand beyond the single layer of ghost cells, thus leading to an incorrect parallel FE solver. To make things worse, *the current literature clearly fails to explain when and why the parallel algorithms and data structures required to support generic conforming FE discretizations atop tree-based adaptive meshes are correct; see, e.g., [14, Sect. 3.3.4]*.

The main contribution of the paper is to address *in a rigorous way* the development of algorithms and data structures for parallel adaptive FE analysis on tree-based meshes endowed with SFCs. These are part of a generic FE simulation framework *that is designed to be general enough to support a range of cell topologies, recursive refinement rules for the generation of adaptive trees endowed with SFCs, and various types of conforming FEs*. Starting with a set of (reasonable) assumptions on the basic building blocks that drive the generation of the forest-of-trees, namely, a recursive refinement rule and a SFC index [1], we infer results based on mathematical propositions and proofs, yielding the (correctness of the) parallel algorithms in our framework.

The framework follows a *two-layered meshing approach*, namely an inner, light-weight layer encoding the forest-of-trees, handled by an external specialized meshing engine, and an outer representation of the adaptive mesh suitable for the implementation of generic adaptive FE spaces. The approach of reconstructing a rich mesh data structure to support generic FEs from a specialized forest-of-trees meshing engine is not new in itself. The excellent work in [14] discusses the particular approach followed by the state-of-the-art `deal.II` FE software library [15] in order to support generic adaptive FE spaces atop parallel forest-of-octrees meshes. However, *both the outer layer mesh representation itself, and the approach followed in this work to reconstruct the outer layer from the inner one are new*. In particular, [14] follows the so-called match-tree-recursive approach, while the one here is based on local neighborhood information across the cell boundaries in the adapted mesh.

Atop this two-layered mesh representation, we design the rest of algorithms and data structures in the framework towards the final assembly of the discrete system of linear equations. In order to be able to leverage non-overlapping domain decomposition solvers, the algorithms in the framework are designed assuming that one deals with a subassembled data layout for the linear algebra data structures on the interface among subdomains.[1] In any case, by means of an extra final step, which is also discussed in the paper, one can construct a global DOF numbering across the whole domain, and thus support fully-assembled global matrices and preconditioners, such as parallel Algebraic MultiGrid (AMG), in the same framework. We can in particular leverage parallel distributed-memory linear algebra software packages, such as, e.g., `PETSc` [18], or `TRILINOS` [19],

An implementation of the framework is available at `FEMPAR` [20], an open source Object-Oriented (OO) Fortran200X scientific software package for the High Performance Computing (HPC) simulation of complex multiphysics problems governed by PDEs at large scales. `FEMPAR` uses `p4est` as its specialized forest-of-octrees meshing engine, although any other engine able to fulfill a set of requirements described in the paper could be plugged in the algorithms of the framework as well. A strong scaling study of this implementation when applied to Poisson and Maxwell problems reveals remarkable scalability up to 32.2K CPU cores and 482.2M degrees of freedom. Besides, a comparison of `FEMPAR` performance with that of `deal.II`, reveals at least competitive performance, and at most factor 2-3 improvements on a massively parallel supercomputer.

---

[1]We refer the reader to [16, 17] for the expected performance and scalability of the domain decomposition solver within `FEMPAR`.

This paper is structured as follows. In Sect. 2, we overview tree-based AMR with SFCs. In Sect. 3, we present the outer mesh layer in our two-layered mesh approach. In Sect. 4, we discuss the construction of conforming FE spaces *in parallel*, and that of sub- and fully-assembled linear systems. The main contributions of the paper are concentrated in these two latter sections. When appropriate, differences among our approach and the one in [14] are highlighted for context. In Sect. 5, we assess the parallel strong scalability and performance of the proposed algorithms and data structures as implemented in `FEMPAR` for the Poisson and Maxwell PDEs on the MN-IV petascale supercomputer, and compare them with their counterparts in `deal.II` for the former problem. Finally, we draw conclusions in Sect. 6. In order to increase the readability of the paper, the proofs of all propositions in the paper are deferred to Appendix A. Those propositions which yield intermediate results and are only required to prove other propositions are also stated in Appendix A.

## 2. Overview of tree-based AMR endowed with space-filling curves

We describe the so-called tree-based AMR with SFCs approach for scalable mesh generation and partitioning. The idea was originally proposed in [2] for the particular case of octrees, and extended to general trees in [1]. Let $\Omega \subset \mathbb{R}^d$ be an open bounded domain in which our PDE problem is posed. AMR with SFCs can be seen as a two-level decomposition of $\Omega$, referred to as macro and micro level, resp.[2] In the macro level, we assume that there is a partition $C$ of $\Omega$ into cells $K \in C$ such that each of these cells can be expressed as a differentiable homeomorphism $\Phi_K$ over a set of admissible reference polytopes [20]. The mesh $C$ is referred to as the coarse mesh and it is assumed to be a *conforming mesh* (see Sect. 2.1). In the micro level, each of the cells of $C$ becomes the root of an adaptive tree. This two-tier adaptive structure is referred to as *forest-of-trees*. It represents a locally refined mesh $\mathcal{T}$. The process that generates it is essentially characterized by defining two complementary ingredients, namely a (recursive) refinement rule (see Sect. 2.2) and an SFC index (see Sect. 2.3). We note that the ideas presented in this section are not new. However, we cover them to an extent such that we can precisely state a set of assumptions on the aforementioned two ingredients, and define the notation that we require to give mathematical rigor to the derivation of the algorithms in our framework.

2.1. **Polytopes and coarse mesh.** Reference polytopes for FE analysis are usually cubes or tetrahedra (and their extensions to arbitrary dimensions), but prisms and pyramids can also be used. A reference polytope spans an open domain. We consider a partition of its boundary into *disjoint* polytopes of lower dimension (i.e., vertices, edges without endpoints, etc.), denoted as $n$-faces, with $0 \le n < d$ being their topological dimension; the $d$-face is the polytope itself. Furthermore, the $n$-faces on the boundary of an $m$-face, $n < m$, of $T$ are also $n$-faces of $T$. For example, with $d = 3$, the boundary is composed of VEFs, representing the 0, 1, and 2-faces of the cell, resp. Hereafter, we may abuse notation by using the acronym VEF for any $n$-face with $n \in [0, d)$. The set of VEFs of $T$ is represented with $\mathcal{F}_T$.

Cells in the physical domain $K \in \mathcal{T}$ are determined by $\Phi_K$ and the reference polytope $T$. The map $\Phi_K$ applied to every $t \in \mathcal{F}_T$ generates the set of *local* VEFs of $K$, denoted as $\mathcal{F}_K$. A local VEF can be understood as a tuple of the cell $K$ it belongs to and the geometrical domain it represents, which we denote as $[f]$. Global VEFs only represent the geometrical domain. Thus, given a local VEF $f$, its corresponding global VEF is $[f]$. Formally, local VEFs are glued together into global VEFs as dictated by an equivalence relation $\sim$ such that $f \sim g$ iff $[f] = [g]$. We represent with $\mathcal{F}$ the set of *global* VEFs of $\mathcal{T}$, i.e., the quotient space of all equivalence classes (global VEFs) resulting from the application of $\sim$ to the elements of $\bigcup_{K \in \mathcal{T}} \mathcal{F}_K$. $[\cdot]$ is the so-called local-to-global VEF map.

As mentioned at the beginning of Sect. 2, in AMR with SFCs, the coarse mesh $C$ is assumed to be conforming. A formal definition of this property is as follows.

**Definition 2.1** (Conforming mesh). *A mesh $\mathcal{T}$ is conforming if, for any two cells $K, K' \in \mathcal{T}$ with $\overline{K} \cap \overline{K'} \ne \emptyset$, there exist $f \in \mathcal{F}_K$, $f' \in \mathcal{F}_{K'}$ such that $[\overline{f}] = [\overline{f'}] = \overline{K} \cap \overline{K'}$.*

---

[2]We note that this two-level construction is for geometrical purposes only.

2.2. **Refinement rule.** Let us consider a cell $K \in \mathcal{C}$ in the coarse mesh of $\Omega$. A *refinement rule* prescribes how one may subdivide the coarse cell $K$ into finer cells that cover the same region as $K$. $K$ is referred to as the *parent* cell, and the latter ones, (its) *children* cells. The refinement rule is usually defined at the polytope and then mapped to the physical space using $\Phi_K$. The reverse application of the refinement rule, i.e., the replacement of the children cells by its parent, is referred to as coarsening. Mesh generation in this context is a hierarchical process based on the recursive application of refinement and coarsening. At each level in the hierarchy, some cells are marked for refinement, and some other for coarsening. A cell marked for refinement is replaced by its children cells following the refinement rule. On the other hand, if all children cells of a given parent are marked for coarsening, they are collapsed into the parent cell. As a result of this process for all cells in $\mathcal{C}$, we obtain a *forest* of tree-like refinement structures, with a tree rooted at every $K \in \mathcal{C}$. We make the following assumption on the refinement rule.

**Assumption 2.2** (Admissible refinement rule). *The cell refinement rule over a cell $K$ provides a conforming mesh $\mathcal{R}_K$ of $K$ (into children cells) and an orientation to the n-faces of $\mathcal{R}_K$. The restriction of $\mathcal{R}_K$ to any n-face $f \in \mathcal{F}_K$ with $n > 0$ is a non-trivial partition of $f$, i.e., these n-faces are subdivided. Besides, when recursively applying the refinement rule to a cell, the resulting mesh is non-degenerate: there is a constant $\rho > 0$ independent of the number of levels of refinement such that, for any cell $K$ in the resulting mesh, there is a ball of diameter $\rho \operatorname{diam}(K)$ contained in $K$, where $\operatorname{diam}(K)$ denotes the diameter of $K$.*

We need Ass. 2.2 in order to be able to mathematically prove the correctness of the algorithms in our framework. In practice, this assumption is not restrictive, since it holds for those refinement rules and SFCs which are at the heart of the state-of-the-art software in tree-based adaptive meshes endowed with SFCs [1, 2], i.e., uniform refinement rules of hexahedra, tetrahedra, prisms, and pyramids (and 2D counterparts).

We define the following relation between the VEFs of $\mathcal{R}_K$, represented with $\mathcal{F}_{\mathcal{R}_K} \doteq \bigcup_{K_c \in \mathcal{R}_K} \mathcal{F}_{K_c}$, and the n-faces in $\mathcal{F}_K \cup \{K\}$.

**Definition 2.3** (Owner of refined VEF). *For every VEF $f \in \mathcal{F}_{\mathcal{R}_K}$, we define its owner n-face $O_f$ as the unique n-face in $\mathcal{F}_K \cup \{K\}$ that contains it, i.e., $f \subset O_f$.*[3]

By Def. 2.3, $O_f$ has a topological dimension greater than or equal to the one of $f$.

A particular adaptive mesh $\mathcal{T}$ of $\Omega$ to be used for FE discretization is defined as the union of all leaf cells (i.e., cells with no children) in the forest. We denote by $|\mathcal{T}|$ the number of cells in $\mathcal{T}$. For every cell $K \in \mathcal{T}$, we can define $\ell(K)$ as the *level of refinement* of $K$ in the forest, with $\ell(K) = 0$ if $K \in \mathcal{C}$ and $\ell(K) > 0$ otherwise. If, for any two cells $K, K' \in \mathcal{T}$, we have that $\ell(K) > \ell(K')$, then $K$ is *finer* than $K'$, and $K'$ is *coarser* than $K$.

2.3. **The SFC index.** Let us assume that we bound the maximum level of refinement for any forest-of-trees that can be built by means of the hierarchical process described in Sect. 2.2; this is the case in practice since available memory is limited. Let us denote with $n > 0$ such maximum level of refinement, and with $\mathcal{S}_n$ the refinement tree that results from $n$ recursive applications of the refinement rule to all leaves. $\mathcal{S}_n$ includes the set of all cells with a maximum refinement level of $n$ that can be potentially constructed from $\mathcal{C}$ by means of AMR. With this notation, we can readily introduce the concept of *SFC index*.[4] The maximum level subscript is omitted in the definition.

**Definition 2.4.** *An SFC index $\mathcal{I}$ on $\mathcal{S}$ is formally defined as a map $\mathcal{I} : \mathcal{S} \to \mathbb{N}_0$ that fulfills the following three properties. First, the map $\mathcal{I} \times \ell : \mathcal{S} \to \mathbb{N}_0 \times \mathbb{N}_0$ is injective, and thus the pair composed by the SFC index and the refinement level of a cell uniquely identifies it among all cells in $\mathcal{S}$. Second, for any $K, K' \in \mathcal{S}$, where $K'$ is a descendant of $K$, then $\mathcal{I}(K) \leq \mathcal{I}(K')$. Additionally, given a $K''$ that is not a descendant of $K$ and $\mathcal{I}(K) < \mathcal{I}(K'')$, it holds that $\mathcal{I}(K') < \mathcal{I}(K'')$.*

A key requirement from the computational viewpoint is to choose the $\mathcal{I}$ mapping such that functions that locally operate on a cell (or small set of cells) take *constant time*, independently of the level of refinement of the cell. Local cell functions include computing the SFC index of a cell, determining the SFC index of its parent, children, and neighbors with the same refinement level, or computing vertex coordinates. Besides, the exploitation of the SFC index for implementing such operations has to result in significant memory savings with respect to unstructured meshing, in which a list of neighbors has to be stored, as well as the coordinates of all vertices in the mesh, for each cell. An SFC index that fulfills Def. 2.4, and the aforementioned requirement, is highly suited for efficient data storage and traversal of the adaptive mesh, fast computation of hierarchy and neighborship relations among the mesh cells, and scalable partitioning and dynamic load balancing [1].[5]

---

[3]Along the paper we will use $\subset$ (resp., $\supset$) to denote subset (resp., superset) inclusion. The proper (a.k.a. strict) variant of this operator will be denoted as $\subsetneq$ (resp., $\supsetneq$).

[4]The definition in the sequel has been adapted from [1], where a novel approach to the theory of discrete SFCs suitable for tree-based AMR is presented.

2.4. **Non-conformity and cell neighbors.** Tree-based meshes provide multi-resolution capability by local adaptation, i.e., cells in $K \in \mathcal{T}$ might have different value for $\ell(K)$. However, these meshes are (potentially) *non-conforming*, i.e., they contain the so-called *hanging VEFs*. These occur at the interface of neighboring cells with different refinement levels.

In order to provide support to FE applications, tree-based AMR with SFCs engines also require to keep track of neighboring relationships between cells in $\mathcal{T}$ (apart from hierarchical relationships). One possible way of describing these is by means of the notion of *cell neighbors across local n-faces*, which we define in the sequel.

**Definition 2.5** (Neighbors of a cell across local *n*-faces). *The neighbors of $K \in \mathcal{T}$ across its n-face $f \in \mathcal{F}_K$ are classified into the sets $\mathcal{T}_{K,f}$, $\mathcal{T}^+_{K,f}$, and $\mathcal{T}^-_{K,f}$, referred to as conformal, higher-level, and lower-level neighbors, resp. They are composed of those cells $K' \in \mathcal{T} \setminus K$ that contain $f' \in \mathcal{F}_{K'}$ such that: (a) $[f] \odot [f']$, with $\odot$ being $=$, $\supsetneq$, and $\subsetneq$ for $\mathcal{T}_{K,f}$, $\mathcal{T}^+_{K,f}$, and $\mathcal{T}^-_{K,f}$, resp.; (b) $[\overline{f}] \cap [\overline{f'}] = \overline{K} \cap \overline{K'}$.*

The connectivity information underlying Def. 2.5 is required by our framework in order to (re)build a mesh data structure suitable for the construction of generic conforming FEs; see Sect. 3. We expect any forest-of-trees mesh engine to be able to provide this sort of relationships, as these are indeed internally determined as a requirement of some of the algorithms provided by the engine; see, e.g., Balance or Ghost in Sect. 2.6.

With the cell neighbors across *n*-faces, one can compute the local-to-global VEF map $[\cdot]$; see Alg. 1 for more details. Given $f \in \mathcal{F}_K$ and $F = [f]$, we define $\mathcal{T}_{K,F} \doteq \mathcal{T}_{K,f}$ (analogously for $\mathcal{T}^+_{K,F}$ and $\mathcal{T}^-_{K,F}$). We denote by $[\mathcal{F}_K] \doteq \{[f] : f \in \mathcal{F}_K\}$ the set of global VEFs of $K$. We next introduce a proposition on $\mathcal{T}_{K,F}$, $\mathcal{T}^+_{K,F}$, and $\mathcal{T}^-_{K,F}$. In order to prove this proposition (and some of those in Sect. 3), we need the following assumption on the recursive refinement procedure.

**Assumption 2.6** (Uniform refinement conformity). *The mesh composed of all leaves of the refinement tree $S_\ell$ obtained after an arbitrary level $\ell$ of uniform refinements of $C$ is also conforming.*

The uniform refinement conformity in Ass. 2.6 requires a refinement rule consistent among cells in the coarse mesh. We note that this assumption is already fulfilled by the standard uniform refinement rules mentioned after Ass. 2.2.

**Proposition 2.7.** *Given a cell $K \in \mathcal{T}$, and an n-face $F \in [\mathcal{F}_K]$ such that $n > 0$, then $\mathcal{T}_{K,F}$ can only be composed of neighbor cells at the same refinement level as $K$. On the other hand, $\mathcal{T}^-_{K,F}$ (resp., $\mathcal{T}^+_{K,F}$) can only be composed of neighbor cells with lower (resp., higher) refinement level than that of $K$.*

However, for $n = 0$, $\mathcal{T}_{K,F}$ might be composed of neighbors at the same, higher, or lower-level of refinement than $K$. This can be readily observed, e.g., in Fig. 1a, for any corner that meets at the boundary of cells with different levels of refinement.

2.5. **Balanced forest-of-trees.** For FE applications, mesh non-conformity makes harder the construction of conforming FE spaces, and the subsequent steps in the simulation; see Sect. 4. However, common practice in order to significantly alleviate this extra complexity consists of enforcing the so-called 2:1 balance constraint (a.k.a. balance or mesh regularity condition). A general definition of the 2:1 balance condition, parameterized by an integer $0 \leq k < d$, is as follows.

**Definition 2.8.** *A forest-of-trees mesh $\mathcal{T}$ is 2:1 k-balanced if and only if, for any cell $K \in \mathcal{T}$, and n-face $F \in [\mathcal{F}_K]$, with $n \in [k, d)$, there is no neighbor $K'$ of $K$ across $F$ (see Def. 2.5) such that $|\ell(K') - \ell(K)| > 1$.*

In other words, geometrically neighboring cells may differ at most by a single level of refinement, with the notion of cell neighborhood depending on the value of $k$. Fig. 1a illustrates a forest-of-quadtrees, with two quadtrees (i.e., $|C| = 2$), which is 2:1 0-balanced, i.e., balance across corners and facets. We note that, as a consequence of Prop. 2.7, and Def. 2.8, the $\mathcal{T}^+_{K,f}$, and $\mathcal{T}^-_{K,f}$ sets for local *n*-faces $f$ with $n \geq \max(k, 1)$, are composed of cells $K'$ such that $\ell(K') = \ell(K)+1$ and $\ell(K') = \ell(K)-1$, resp. On the other hand, if $n = k = 0$, then $\mathcal{T}_{K,f}$ may be composed of neighbors at the same, one unit higher, or one unit lower level of refinement than $K$; see Fig. 1a.

---

[5]Dynamic load balancing is the ability of an adaptive mesh to be re-distributed in the presence of an unacceptable amount of load imbalance, e.g., the one generated by means of AMR in a highly localized region of $\Omega$.

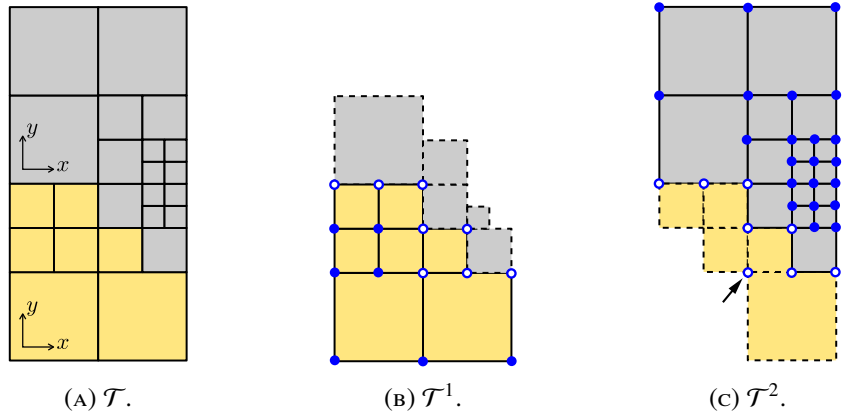

(A) $\mathcal{T}$. (B) $\mathcal{T}^1$. (C) $\mathcal{T}^2$.

FIGURE 1. 2:1 0-balanced forest-of-quadtrees mesh with two quadtrees (i.e., $|C| = 2$) distributed (non-uniformly) among two processors (i.e., $P = 2$) with 0-ghost cells, 1:4 refinement and the Morton SFC index [2]. Cells in $\mathcal{T}_L^p$ are depicted with continuous boundary lines, while those in the 0-ghost layer $\mathcal{T}_G^p$ with dashed ones. All vertices in $\mathcal{F}_L^p$ are depicted with circles. In particular, vertices in $\mathcal{F}_I^p$ are depicted with blue circles, whereas vertices in $\mathcal{F}_\Gamma^p$ are depicted with unfilled blue circles. The mesh vertex in $\mathcal{T}^2$ pointed by the arrow is in $\mathcal{F}_L^p$ as it fulfills Def. 3.9 (d).

In Sect. 4, it will become clear why this constraint extremely simplifies the construction of conforming FE spaces, especially in a distributed-memory context.

2.6. **Forest-of-trees handler operations.** In practice, parallel tree-based AMR using SFCs is a specialized feature that numerical applications typically outsource to an external adaptive meshing engine. In the interface among these, a set of core application-level operations have been identified as cornerstone in order to fully realize this functionality in numerical applications. These are briefly outlined in the sequel. (a) `New`: creates a new uniformly refined forest, up to a user-provided level, from a data structure describing the connectivity of mesh cells in $C$; (b) `Adapt`: refines and coarsens the current forest according to a user-provided criterion; (c) `Partition`: redistributes the forest leaves among processors for dynamic load balancing; (d) `Balance`: ensures the 2:1 $k$-balance condition among neighboring cells by local refinement where required; (e) `Ghost`: creates a data structure that contains the so called $s$-ghost cell set, i.e., a layer of off-processor leaf cells that are neighbors of cells in the current processor (see Sect. 3); (f) `Iterate`: given a 2:1 $k$-balanced forest, provides a mechanism to iterate over its leaf cells, and *a subset of* the inter-cell interfaces (i.e., $n$-faces such that $n \geq k$), while letting applications get local neighborhood information of each interface visited, including both conforming and non-conforming cell interfaces. We refer to [1–3] and references therein for a comprehensive presentation of the algorithms behind these operations, implementation details, and cost and communication volume analysis, among others.

3. A GENERAL FE-SUITABLE DISTRIBUTED ADAPTIVE MESH REPRESENTATION

In this section we present a distributed adaptive mesh representation that supports generic FE spaces built atop. This data structure implements the outer mesh layer in our two-layered framework; see Sect. 1. In Sect. 3.1 and 3.2, we formally define the sort of data it handles (e.g., adjacencies among cells and global VEFs). The exposition is supplemented with a set of propositions that are required in order to prove the correctness of the algorithms presented in Sect. 3.3 and Sect. 4. The concepts underlying our adaptive mesh data structure are not tailored to a particular tree-based AMR with SFCs technology or cell topology.

3.1. **Cells and global VEFs adjacencies.** Our adaptive mesh representation considers three different kinds of adjacency relations among cells and global VEFs. First, given a cell $K \in \mathcal{T}$, $[\mathcal{F}_K]$ keeps track of the global VEFs (equivalence classes) corresponding to the local VEFs of $K$. The remaining two relations are neighborship relations. Given a global VEF $F \in \mathcal{F}$, the set of cells around the VEF is defined as $\mathcal{T}_F \doteq \{K \in \mathcal{T} : F \in [\mathcal{F}_K]\}$. For our purposes, the definition of the *coarser cells around a global VEF* is essential.

**Definition 3.1** (Coarser cells around a global VEF). *The set of coarser cells around a VEF $F \in \mathcal{F}$ is represented with $\tilde{\mathcal{T}}_F$ and is composed by the cells $K \in \mathcal{T}$ such that : (a) $F \notin [\mathcal{F}_K]$; (b) $F \subset \overline{K}$.*

The $\mathcal{T}_F$ and $\tilde{\mathcal{T}}_F$ sets can be readily obtained from the neighbors of cells across local $n$-faces, as stated in the following two propositions. For conciseness, we define the sets

$$\mathcal{S}_{K,F} \doteq \bigcup_{G \in [\mathcal{F}_K]: F \subset \overline{G}} \mathcal{T}_{K,G}, \quad \mathcal{S}_{K,F}^- \doteq \bigcup_{G \in [\mathcal{F}_K]: F \subset \overline{G}} \mathcal{T}_{K,G}^-, \quad \mathcal{S}_{K,F}^+ \doteq \left\{ K' \in \left( \bigcup_{G \in [\mathcal{F}_K]: F \subset \overline{G}} \mathcal{T}_{K,G}^+ \right) : F \in [\mathcal{F}_{K'}] \right\}.$$

**Proposition 3.2.** *Given a cell $K \in \mathcal{T}$ and $F \in [\mathcal{F}_K]$, then we have that:*

$$\mathcal{T}_F = \{K\} \cup \mathcal{S}_{K,F} \cup \mathcal{S}_{K,F}^+ \cup \{K' \in \mathcal{S}_{K,F}^- : F \in [\mathcal{F}_{K'}]\}, \tag{1}$$

$$\tilde{\mathcal{T}}_F = \{K' \in \mathcal{S}_{K,F}^- : F \notin [\mathcal{F}_{K'}]\}. \tag{2}$$

**Proposition 3.3.** *Given an $n$-face $F \in \mathcal{F}$, with $n > 0$, and a cell $K \in \mathcal{T}$ such that $F \in [\mathcal{F}_K]$, then we have that: (a) $\mathcal{T}_F = \mathcal{S}_{K,F} \cup \{K\}$; and (b) $\tilde{\mathcal{T}}_F = \mathcal{S}_{K,F}^-$.*

The set $\tilde{\mathcal{T}}_F$ can be used to classify $\mathcal{F}$ into *regular* and *hanging* VEFs.

**Definition 3.4** (Regular and hanging VEFs). *A VEF is regular if $\tilde{\mathcal{T}}_F = \emptyset$ and hanging otherwise.*

We denote as $\mathcal{F}_R$ and $\mathcal{F}_H$ the set of regular and hanging VEFs, resp. Clearly, $\{\mathcal{F}_R, \mathcal{F}_H\}$ is a partition of $\mathcal{F}$.

**Proposition 3.5.** *Given a 2:1 $k$-balanced mesh $\mathcal{T}$ and $F \in \mathcal{F}_H$ such that its owner VEF $O_F$ is an $n$-face with $n \geq k$, then the sets $\tilde{\mathcal{T}}_F$ and $\mathcal{T}_{O_F}$ are identical (or, equivalently, $O_F \in \mathcal{F}$).*

The following *key property* holds for hanging VEFs.

**Proposition 3.6.** *Given a 2:1 $k$-balanced forest-of-trees mesh $\mathcal{T}$ and $F \in \mathcal{F}_H$ such that $O_F$ is an $n$-face with $n \geq k$, it holds that $O_F \in \mathcal{F}_R$. Besides, any $n$-face $E$ of $\overline{O_F}$ with $n \geq k$ is also regular, i.e., $E \in \mathcal{F}_R$.*

As a consequence of Prop. 3.6, one enjoys the following two cornerstone benefits: (a) *only* single-level (a.k.a. direct) hanging node constraints are required when constructing global conforming FE spaces; (b) in a distributed-memory context, all such constraints can be resolved locally with a single layer of ghost cells. This will be revisited in Sect. 4. A stronger result is proved below for 2:1 1-balanced forest-of-trees.

**Proposition 3.7.** *Given a 2:1 1-balanced forest-of-trees mesh $\mathcal{T}$ and $F \in \mathcal{F}_H$, then $O_F \in \mathcal{F}_R$. Besides, any $n$-face $E$ of $\overline{O_F}$ is also regular, i.e., $E \in \mathcal{F}_R$.*

Prop. 3.7 implies that, for those global conforming FE spaces for which vertices carry out DOFs (e.g., those constructed from Lagrangian FEs), 2:1 0-balanced trees are not required to have (a) and (b) above, only 2:1 1-balance is required. Provided this is acceptable for the problem at hand from a numerical point of view, 2:1 1-balance may lead to computational savings, as enforcing it requires less balance refinement than 0-balance in general. While the result itself is known [21], up to the authors' knowledge, it has not been stated formally nor mathematically proven.

3.2. **Distributed-memory context.** With the definitions associated to the global adaptive mesh $\mathcal{T}$ so far, we can readily discuss *its representation in a parallel distributed-memory context*. All cells $K \in \mathcal{T}$ are assigned *a processor owner* $p = 1, \ldots, P$, with $P$ being the number of processors involved in the parallel computation. Given this cell-to-processor ownership mapping, the local portion of the global mesh $\mathcal{T}$ that processor $p$ stores locally, referred to as $\mathcal{T}^p \subset \mathcal{T}$, is defined as the union of two *disjoint* sets of cells, i.e., $\mathcal{T}^p \doteq \mathcal{T}_L^p \cup \mathcal{T}_G^p$, $\mathcal{T}_L^p \cap \mathcal{T}_G^p = \emptyset$, with $\mathcal{T}_L^p$ being the *local* cells set, and $\mathcal{T}_G^p$ the so-called *ghost* cells set. The former includes those cells that processor $p$ owns. By construction, the set $\{\mathcal{T}_L^p\}$, for $p = 1, \ldots, P$, is a partition of $\mathcal{T}$. The formal definition of the *ghost* cells set is also parameterized by a parameter $s$, with $0 \leq s < d$.

**Definition 3.8.** *The $s$-ghost cell set includes all cells $K \in \mathcal{T} \setminus \mathcal{T}_L^p$ (i.e., off-processor cells) that are neighbors of cells in $\mathcal{T}_L^p$ across $n$-faces with $n \geq s$ (see Def. 2.5).*

The sets $\mathcal{T}_L^p$ and $\mathcal{T}_G^p$ with $s = 0$, for $p = 1, 2$, are illustrated in Fig. 1b, 1c, resp., for the forest-of-quadtrees in Fig. 1a. We note that a given cell in $\mathcal{T}$ belongs to only one $\mathcal{T}_L^p$ set. It, however, belongs to either none, one, or more $\mathcal{T}_G^p$ sets.

Since a given processor only stores a local portion of the global mesh, i.e., $\mathcal{T}^p$, it can only store a subset of the global VEFs set $\mathcal{F}$, denoted as $\mathcal{F}^p \subset \mathcal{F}$, and referred to as the proc-local VEFs set. $\mathcal{F}^p$ is generated by gluing together the local VEFs that lie on the boundary of (the local and ghost) cells in $\mathcal{T}^p$. The sets $\mathcal{T}_{K,f}$, $\mathcal{T}_{K,f}^-$, and $\mathcal{T}_{K,f}^+$ are also restricted to cells in $\mathcal{T}^p$. We denote their restriction as $\mathcal{T}_{K,f}^p \doteq \mathcal{T}_{K,f} \cap \mathcal{T}^p$, $\mathcal{T}_{K,f}^{p,+} \doteq \mathcal{T}^p \cap \mathcal{T}_{K,f}^+$, and $\mathcal{T}_{K,f}^{p,-} \doteq \mathcal{T}^p \cap \mathcal{T}_{K,f}^-$, resp., for all $K \in \mathcal{T}^p$. It turns out that, for all cells $K \in \mathcal{T}_L^p$, and local n-faces $f \in \mathcal{F}_K$ such that $n \geq s$, the processor-local and global sets are equivalent by Def. 3.8. Likewise, $\mathcal{T}_F$ and $\tilde{\mathcal{T}}_F$ are restricted to the cells in $\mathcal{T}^p$. For a given proc-local VEF $F \in \mathcal{F}^p$, we denote their restriction as $\mathcal{T}_F^p \doteq \mathcal{T}_F \cap \mathcal{T}^p$ and $\tilde{\mathcal{T}}_F^p \doteq \tilde{\mathcal{T}}_F \cap \mathcal{T}^p$, resp. By the equivalences in Props. 3.2, 3.3, we have that $\mathcal{T}_F^p = \mathcal{T}_F$ and $\tilde{\mathcal{T}}_F^p = \tilde{\mathcal{T}}_F$ for those n-faces $F \in \mathcal{F}^p$ such that $n \geq s$, and there is at least one local cell in $\mathcal{T}_F^p$. As shown along the paper, this equivalence has key implications.

The set of proc-local VEFs $\mathcal{F}^p$, as in the case of cells, can be split into two disjoint sets of VEFs, i.e., $\mathcal{F}^p \doteq \mathcal{F}_L^p \cup \mathcal{F}_G^p$. This separation into the $\mathcal{F}_L^p$ and $\mathcal{F}_G^p$ sets is, however, of different nature compared to that of the cells.

**Definition 3.9** ($\mathcal{F}_L^p$ and $\mathcal{F}_G^p$ sets). *A VEF $F \in \mathcal{F}^p$ is in $\mathcal{F}_L^p$ if it fulfills one of the four following conditions: (a) at least one of the cells in $\mathcal{T}_F^p$ is local, i.e., $\mathcal{T}_F^p \cap \mathcal{T}_L^p \neq \emptyset$; (b) all cells in $\mathcal{T}_F^p$ are ghost but at least one cell in $\tilde{\mathcal{T}}_F^p$ is local, i.e., $\tilde{\mathcal{T}}_F^p \cap \mathcal{T}_L^p \neq \emptyset$; (c) all cells in $\mathcal{T}_F^p$ are ghost and there exists $F' \in \mathcal{F}_L^p$ such that $F = O_{F'}$; (d) all cells in $\mathcal{T}_F^p$ are ghost and $F$ lies at the boundary of a VEF that fulfills (c). It is in $\mathcal{F}_G^p$ otherwise.*

Thus, a global VEF $F \in \mathcal{F}$ might be in $\mathcal{F}_L^p$ at multiple processors, while a local cell cannot. We will refer to this kind of VEFs as *interface* VEFs, while we will use the term *interior* to refer to those VEFs which are in $\mathcal{F}_L^p$ only at a single processor. We denote the former and latter sets as $\mathcal{F}_\Gamma^p$ and $\mathcal{F}_I^p$, resp. Clearly, $\{\mathcal{F}_I^p, \mathcal{F}_\Gamma^p\}$ is a partition of $\mathcal{F}_L^p$. Those vertices in the sets $\mathcal{F}_L^p$, $\mathcal{F}_I^p$, and $\mathcal{F}_\Gamma^p$ for the forest-of-quadtrees in Fig. 1a distributed among two processors, are illustrated in Fig. 1b, 1c for $p = 1, 2$, resp. It remains to define a classification of VEFs of $\mathcal{T}^p$ into sets of regular and hanging VEFs suitable in a distributed context, i.e., that only requires processor-local information. With this aim, we define the concept of proc-regular and proc-hanging VEFs.

**Definition 3.10** (Proc-regular and proc-hanging VEFs). *A VEF $F \in \mathcal{F}^p$ is proc-regular if $\tilde{\mathcal{T}}_F^p = \emptyset$, and proc-hanging otherwise.*

The set of proc-regular and proc-hanging VEFs are denoted as $\mathcal{F}_R^p$ and $\mathcal{F}_H^p$, resp. We stress that, for VEFs in $\mathcal{F}_G^p$, this definition of proc-regular (resp., proc-hanging) VEFs is not equivalent to regular (resp., hanging) VEFs. In Fig. 2a, 2b, 2c, 2d, we illustrate the $\mathcal{F}_R^p$ and $\mathcal{F}_H^p$ sets, with $p = 1, 2$, resp., for the forest-of-quadtrees in Fig. 1a. The ghost vertex and ghost face pointed by an arrow in Fig. 2c and 2d are such that they belong to $\mathcal{F}_H$ in $\mathcal{T}$ but to $\mathcal{F}_R^p$ in $\mathcal{T}^p$. Fortunately, the algorithms that run on top of the mesh *only* require these definitions to be equivalent for n-faces in $\mathcal{F}_L^p$, where the values of n are again determined by the FE space at hand. This invariant holds, as stated in the next proposition.

**Proposition 3.11.** *For a 2:1 k-balanced forest-of-trees mesh $\mathcal{T}$ with k-ghost cells, then $\mathcal{F}_R^p \cap \mathcal{F}_L^p = \mathcal{F}_R \cap \mathcal{F}_L^p$ and $\mathcal{F}_H^p \cap \mathcal{F}_L^p = \mathcal{F}_H \cap \mathcal{F}_L^p$ for n-faces F with $n \geq k$. This statement also holds for 0-faces when $k = 1$ (i.e., 2:1 1-balance and 1-ghost cell set).*

3.3. **Construction of a FE-suitable distributed adaptive mesh data structure.** Alg. 1 shows a simplified pseudocode of the process in charge of building the outer mesh layer in our framework. In a distributed-memory context, each processor $p$ executes its own instance of Alg. 1, *with no communication at all involved among processors*. The algorithm assumes the global forest-of-trees mesh $\mathcal{T}$ to be 2:1 k-balanced, with k being a user-level parameter to be set up according to Prop. 4.1; see Sect. 4.3. The input of Alg. 1 are the proc-local variants of the cell neighbors across local n-faces sets presented in Sect 2.4. From this input, Alg. 1 glues together the local VEFs lying at the boundary of the mesh cells in $\mathcal{T}^p$ that satisfy the condition in Line 4, thus generating: (a) $[\mathcal{F}_K]$, for $K \in \mathcal{T}^p$; (b) the proc-local $\mathcal{F}^p$, $\mathcal{F}_H^p$ and $\mathcal{F}_R^p$ VEFs sets; (c) the proc-local $\mathcal{T}_F^p$ and $\tilde{\mathcal{T}}_F^p$ sets; (d) $O_F$ for all $F \in \mathcal{F}_H^p$. Although omitted from Alg. 1 in order to keep the presentation short, the actual algorithm also builds the $\mathcal{F}_L^p$, $\mathcal{F}_G^p$, $\mathcal{F}_I^p$, and $\mathcal{F}_\Gamma^p$ sets (see Sect. 3.2).

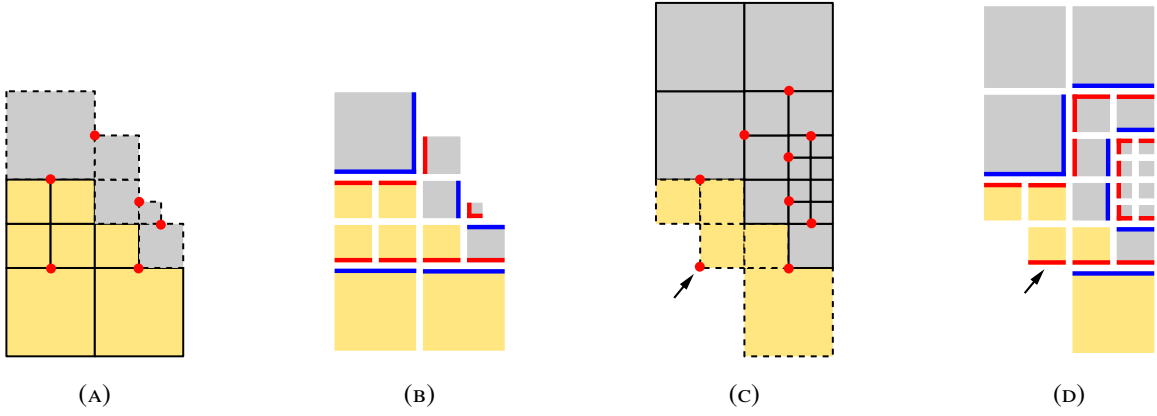

FIGURE 2. Hanging vertices (red circles) and faces (red lines) in $\mathcal{T}^1$ (A and B, resp.) and $\mathcal{T}^2$ (C and D, resp.). Those regular faces that are the owner VEF of a hanging VEF are depicted in blue. The hanging VEFs pointed out by the arrows in $\mathcal{T}^2$ are two examples of ghost VEFs that, while being hanging in $\mathcal{T}$, cannot be identified as such by processor $p = 2$, i.e., they are proc-regular.

The correctness of Alg. 1 is mathematically supported by the propositions in Sect. 3.1 and 3.2. First, due to Prop. 3.2 and 3.3, Alg. 1 is able to obtain $\mathcal{T}_F^p$ from the proc-local variants of the cell neighbors across local $n$-faces sets; see Lines 8-28. Second, Lines 29-34 are well-defined due to Prop. 3.5, 3.6, and 3.7. These lines are in charge of generating the $\tilde{\mathcal{T}}_F^p$ set for a hanging VEF that fulfills the condition in Line 4. To this end, Line 31 triggers the generation of the $\mathcal{T}_F^p$ and $\tilde{\mathcal{T}}_F^p$ sets for its owner VEF. By Prop. 3.5, we know that the owner VEF exists and due to Prop. 3.6 and 3.7, that the depth of the recursive call in Line 31 is always one, as the owner VEF of a hanging VEF is always a regular VEF (i.e., $\tilde{\mathcal{T}}_{O_F}^p = \emptyset$). Besides, due to Prop. 3.5, Line 32 is correct. Third, thanks to Prop. 3.11, it is safe to compute $\mathcal{F}_R^p$ and $\mathcal{F}_H^p$. This can be done at each processor without any inter-processor communication. The only VEFs for which this locally computed mesh information does not coincide with the global one are not required in practice for the parallel generation of global FE spaces and their numerical integration (see Sect. 4).

**Remark 3.12.** *Alg. 1, at several points, has to determine the local VEF $\hat{f}$ of a neighbor cell $\hat{K}$ of $K$ across $g$ that either satisfies $\hat{f} \sim f$ or $\hat{f} \sim O_f$, with $g \in \mathcal{F}_K$ such that $f \subset \bar{g}$. This becomes a requirement that the forest-of-trees inner layer meshing engine has to be able to fulfill.*

**Remark 3.13.** *Alg. 1 follows a fundamentally different approach from its counterpart in* `deal.II` *[14, Fig. 1], and imposes a different set of requirements to the tree-based AMR mesh engine. This latter algorithm matches recursively the forest-of-octrees within the* `deal.II` *mesh data structure and the one in* `p4est`*. In order to do so, it starts from the root octants of the forest, and using a set of queries to* `p4est`*, reconstructs the local part of the mesh as by-product of a full hierarchical AMR process in which the mesh is incrementally transformed by successive refinement and coarsening steps, until the leaves in the former match those in the latter. At each step, the aforementioned set of queries lets* `deal.II` *determine which cells in the current mesh have to be refined and coarsened in the path towards the final match. While this approach takes into account the prior state of the forest-of-octrees within the* `deal.II` *mesh data structure, and thus does not need to actually perform changes in the data structure if no changes have been performed in its* `p4est` *counterpart, it still has to go over a full hierarchical AMR process to determine whether there is a match or not. Besides, when significant changes have been performed in* `p4est`*, it turns to be a quite complex, harder-to-update data structure (i.e., cell hierarchy, n-faces hierarchy, etc.), than the one introduced in this paper. In the experiments in Sect. 5, the implementation of the latter data structure in* `FEMPAR` *turns out to be up to twice faster than the one in* `deal.II`*, even if Alg. 1 (as now conceived) does not try to exploit the previous state of the mesh data structure. We believe that this performance improvement is algorithmic in nature, although we could not fully discard it to be caused by technical differences in the HPC implementation of both libraries.*

**Remark 3.14** (Implementation remark). *Alg. 1 can be implemented for forest-of-octrees using* `p4est` *as the inner layer mesh engine. In particular, this library offers the so-called* `pXest_mesh_t` *data structure, with* `X=4,8` *for $d = 2, 3$, resp., which provides neighbors across vertices, edges, and faces.* `pXest_mesh_t` *is constructed by means of the so-called universal mesh topology iterators, introduced in [3]; see also* `Iterate` *in Sect. 2.6. As a consequence,* `pXest_mesh_t` *only provides neighbors for cells in* $\mathcal{T}_L^p$*. However, Alg. 1 requires these for cells in* $\mathcal{T}_G^p$

**Algorithm 1:** Construction of FE-suitable adaptive mesh from cell neighbors across $n$-faces.

1    $\mathcal{F}_R^p \leftarrow \emptyset; \mathcal{F}_H^p \leftarrow \emptyset$
2    **for** $K \in \mathcal{T}^p$ **do**                                                        /* loop over all cells in the local portion of processor $p$ */
3      **for** $f \in \mathcal{F}_K$ **do**                                                                        /* loop over current cell local VEFs */
4        **if** ($f$ is a 0-face and $k \le 1$) or ($f$ is an $n$-face with $n \ge k$) **then**
5          **if** $[f]$ not created yet **then**                                   /* current VEF first visited from any cell */
6            $[f] \leftarrow F \leftarrow \texttt{new\_proc\_local\_vef}\,()$;                 /* create new VEF equivalence class */
7            $\mathcal{T}_F^p \leftarrow \{K\}; \tilde{\mathcal{T}}_F^p \leftarrow \emptyset$                           /* initialize the $\mathcal{T}_F^p$ and $\tilde{\mathcal{T}}_F^p$ sets */
8            **for** $g \in \mathcal{F}_K : f \subset \bar{g}$ **do**                                    /* Generate $\mathcal{T}_F^p$ (Lines 8-28) */
9              **if** $\mathcal{T}_{K,g}^p \ne \emptyset$ **then**
10                **for** $\hat{K} \in \mathcal{T}_{K,g}^p$ **do**                     /* loop over conformal neighbors of $K$ across $g$ */
11                  $[\hat{f}] \leftarrow F$, with $\hat{f} \in \mathcal{F}_{\hat{K}} : \hat{f} \sim f; \mathcal{T}_F^p \leftarrow \mathcal{T}_F^p \cup \{\hat{K}\}$
12                **end**
13              **end**
14              **if** $\mathcal{T}_{K,g}^{p,-} \ne \emptyset$ **then**
15                **if** $f$ is a 0-face **then**
16                  **for** $\hat{K} \in \mathcal{T}_{K,g}^{p,-} : \exists\, \hat{f} \in \mathcal{F}_{\hat{K}}$ satisfying $\hat{f} \sim f$ **do**
17                    $[\hat{f}] \leftarrow F; \mathcal{T}_F^p \leftarrow \mathcal{T}_F^p \cup \{\hat{K}\}$
18                  **end**
19                **end**
20              **end**
21              **if** $\mathcal{T}_{K,g}^{p,+} \ne \emptyset$ **then**
22                **if** $f$ is a 0-face **then**
23                  **for** $\hat{K} \in \mathcal{T}_{K,g}^{p,+} : \exists\, \hat{f} \in \mathcal{F}_{\hat{K}}$ satisfying $\hat{f} \sim f$ **do**
24                    $[\hat{f}] \leftarrow F; \mathcal{T}_F^p \leftarrow \mathcal{T}_F^p \cup \{\hat{K}\}$
25                  **end**
26                **end**
27              **end**
28            **end**
29            **if** $\{K' \in \mathcal{S}_{K,f}^- : \nexists\, f' \in \mathcal{F}_{K'}, f' \sim f\} \ne \emptyset$ **then**       /* is $f$ hanging? Generate $\tilde{\mathcal{T}}_F^p$ (Lines 29-34) */
30              Let $\hat{K}$ be an arbitrary neighbor cell in $\{K' \in \mathcal{S}_{K,f}^- : \nexists\, f' \in \mathcal{F}_{K'}, f' \sim f\}$ and $\hat{f} \in \mathcal{F}_{\hat{K}}$ such that $\hat{f} \sim O_f$
31              Execute lines 5-36 replacing $f \equiv \hat{f}$ and $K \equiv \hat{K}$
32              $\tilde{\mathcal{T}}_F^p \leftarrow \mathcal{T}_{O_F}^p$
33              $O_F \leftarrow [O_f]$
34            **end**
35            $(\tilde{\mathcal{T}}_F^p = \emptyset)\,?\, \mathcal{F}_R^p \leftarrow \mathcal{F}_R^p \cup \{F\} : \mathcal{F}_H^p \leftarrow \mathcal{F}_H^p \cup \{F\}$ /* "X ? Y : Z" denotes the conditional ternary operator */
36          **end**
37        **end**
38      **end**
39    **end**
40    $\mathcal{F}^p \leftarrow \mathcal{F}_R^p \cup \mathcal{F}_H^p$

*as well. To this end, we developed a nearest neighbors exchange communication stage to complete* `pXest_mesh_t` *prior to the actual execution of Alg. 1. Apart from this, it turns out that, in* `p4est` *v2.2 (latest stable release at the date of writing), the* `p8est_mesh_t` *data structure does not provide the following required information: (a) cell neighbors across edges sets; (b) vertices hanging on a coarse edge across which there are vertex neighbors satisfying Def. 2.5. Fortunately, we could obtain them using edge topology iterators [3].*

4. Handling conforming FE spatial discretizations on non-conforming meshes

4.1. **Problem statement and its conforming FE spatial discretization.** We aim at solving a PDE problem in $\Omega$, supplemented with appropriate boundary conditions on $\partial\Omega$. In order to end up with a computable version of this problem, the FE method requires finite-dimensional spaces of functions $\mathcal{V}_h$ with some approximability properties. A particular type of FE methods, referred to *conforming FE methods*, require $\mathcal{V}_h$ to be *a conforming FE space*, i.e., a subspace of its infinite-dimensional counterpart $\mathcal{V}$, i.e., $\mathcal{V}_h \subset \mathcal{V}$.

In practice, FE spaces are made of functions which are piece-wise polynomial (i.e., smooth) on each cell. Thus, the conformity requirement translates into a trace continuity requirement on the interface among the mesh cells. When the mesh is conforming, the trace continuity requirement for conformity can be achieved combining two different ingredients, that are defined *abstractly* as follows. First, we define the local space of functions $Q(K)$ unisolvent with respect to a set of local DOFs (functionals) $\mathcal{M}_K$ for every cell $K \in \mathcal{T}$.[6] Second, with these cell-wise spaces, the global FE space $\mathcal{V}_h$ is determined by an equivalence class to glue together local DOFs and create the set of global DOFs $\mathcal{N}$. The equivalent class (global DOF) of a local DOF $\alpha$ is represented with $[\alpha]$; $[\cdot]$ is the so-called local-to-global DOF map.

The equivalence relation and the local FE spaces complement each other strategically such that the continuity of global DOF values resulting from gluing together local DOFs, implies the required trace continuity of $\mathcal{V}_h$. Examples of grad-, curl- and div-conforming finite-dimensional spaces are those which are built from Lagrangian, Nédélec (a.k.a. edge), and Raviart-Thomas (a.k.a. face) FEs by imposing trace continuity, and tangent, and normal component trace continuity, resp. Underlying their construction, there exists the notion of global (local) VEF *owner* of a global (local) DOF, i.e., the global (local) VEFs of the triangulation *may carry out* global (local) DOFs. For example, for first-order Lagrangian FEs spaces, only the 0-faces (i.e., vertices) carry out (own) DOFs, while for first-order Nédélec ones, only the 1-faces do. We use $\mathcal{M}_K^f$ to refer to the set of local DOFs owned by $f \in \mathcal{F}_K$, and $\mathcal{N}_F \subset \mathcal{N}$ denotes the set of all equivalence classes (i.e., global DOFs) resulting from the application of the equivalence relation to all cell-local DOFs $\alpha \in \mathcal{M}_K^f$, for all $K \in \mathcal{T}_F$ such that $F = [f]$. For conciseness, we do not cover here how the abstract ingredients are defined for each specific conforming FE space at hand. A comprehensive exposition can be found at Sect. [20, Sect. 3].

4.2. **Generation of proc-local DOFs.** In our distributed-memory framework, each processor restricts itself to the generation of those equivalence classes (i.e., global DOFs) of $\mathcal{N}$ corresponding to $n$-faces (i.e., VEFs and cells) in $\mathcal{T}^p$. Given $F \in \mathcal{F}^p$, we denote by $\mathcal{N}_F^p$ the set of all equivalence classes on $F$ at processor $p$. We define $\mathcal{N}^p \doteq \bigcup_{F \in \{\mathcal{F}^p \cup \mathcal{T}^p\}} \mathcal{N}_F^p$, with $\mathcal{N}^p \subset \mathcal{N}$, and refer to it as the proc-local DOFs set. We also define:

$$\mathcal{N}_R^p \doteq \cup_{F \in \{\mathcal{F}_R^p \cup \mathcal{T}^p\}} \mathcal{N}_F^p, \quad \mathcal{N}_H^p \doteq \cup_{F \in \mathcal{F}_H^p} \mathcal{N}_F^p, \quad \mathcal{N}_L^p \doteq \cup_{F \in \{\mathcal{F}_L^p \cup \mathcal{T}_L^p\}} \mathcal{N}_F^p,$$

$$\mathcal{N}_G^p \doteq \cup_{F \in \{\mathcal{F}_G^p \cup \mathcal{T}_G^p\}} \mathcal{N}_F^p, \quad \mathcal{N}_I^p \doteq \cup_{F \in \{\mathcal{F}_I^p \cup \mathcal{T}_L^p\}} \mathcal{N}_F^p, \quad \mathcal{N}_\Gamma^p \doteq \cup_{F \in \{\mathcal{F}_\Gamma^p\}} \mathcal{N}_F^p.$$

Assuming that one only requires to build a distributed subassembled linear system (see Sect. 4.3), the need for a local-to-global index map $[\cdot]$ can be by-passed by combining a local-to-proc-local index map, denoted hereafter as $[\cdot]_p$, and Remark 4.4. In other words, local DOFs $\alpha$ in $\mathcal{M}_K$, for $K \in \mathcal{T}^p$, are labeled with a proc-local identifier $[\alpha]_p$ in the range $\{1, \ldots, |\mathcal{N}^p|\}$. Thus, one can prevent dealing with 64-bit integers (i.e., global DOF identifiers) across the whole domain) and use 32-bit ones instead (i.e., processor-local DOF identifiers within each subdomain). This reduces memory consumption and bandwidth demands. Furthermore, using proc-local DOF identifiers, to determine whether a given proc-local DOF in $\mathcal{N}^p$ is in a given set, say $\mathcal{N}_H^p$, can be implemented *in constant time*, whereas with global DOF identifiers, one has to pay the cost of a search in a global index set.

4.3. **Hanging DOF constraints. Which are strictly required locally? When/why can they be resolved locally?** Conformity on the interface of cells at different refinement levels must be enforced explicitly by introducing additional linear algebraic constraints in $\mathcal{V}_h$, i.e., the so-called hanging DOF constraints. The structure and set up of such constraints is well-established knowledge; see, e.g., [12] and [22] for grad- and curl-conforming FE spaces, resp.

*In a distributed-memory computing environment, however, it is (much) more intricate.* In general, distributed-memory computers are most efficiently exploited if one maximizes local work while minimizing inter-processor communication. To this end, in this context, we aim to build without communication a local subassembled portion of the global linear system coefficient matrix (and right-hand-side vector). This local portion is of size $\mathcal{N}_R^p \cap \mathcal{N}_L^p$, and is built by means of a FE assembly process restricted to the cells $K \in \mathcal{T}_L^p$ (i.e., local cells).[7] It follows that only the constraints on hanging DOFs touched by local cells have to be resolved in order to build the subassembled portion. In order to be able to set up and apply such constraints locally, without communication, we have to ensure that there is room in $\mathcal{T}^p$ for any of the free DOFs whose values constraint the one of the hanging DOF touched by local cells. In other words, we have to guarantee that the hanging DOF constraints dependencies do not expand beyond the single layer of ghost cells. Prop. 4.1 precisely answers under which conditions this is guaranteed. We observe that the current literature clearly lacks mathematical rigor to address when and why these constraints can be resolved locally; see, e.g., [14, Sect. 3.3.4].

---

[6]Here, we assume the same polynomial space in all cells.

[7]We note that this assembly process has to deal with the application of hanging DOF constraints, following, e.g., the transformation approach in [13]. In a nutshell, this approach carries out the constraint application during FE assembly, in order to directly build the so-called constrained (a.k.a. condensed) linear system as the cell-local matrices and force vectors are built and assembled.

Let us denote with $D(\mathcal{V}_h)$ the dimension of the *n*-face of lowest dimension that has a non-empty set of owned DOFs. For instance, it is 0 for Lagrangian FEs (vertices own DOFs), 1 for edge FEs (edges own DOFs, vertices do not), 2 for face FEs (faces own DOFs, vertices/edges do not), and $d$ (the space dimension) for Discontinuous Galerkin (DG) (only cells own DOFs since no inter-cell continuity must be enforced). We have the following result.

**Proposition 4.1.** *Let us consider a distributed 2:1 k-balanced forest-of-trees mesh $\mathcal{T}$ with k-ghost cells. If $\max(1, D(\mathcal{V}_h)) \geq k$, all hanging DOF constraints are* direct, *i.e., they only depend on regular DOFs. Furthermore, all constraints of DOFs owned by n-faces in $\mathcal{F}_H \cap \mathcal{F}_L^p$ only depend on DOFs owned by $\mathcal{F}_R \cap \mathcal{F}_L^p$, and thus can be solved in parallel with processor-local information.*

In other words, as by-product of the journey to Prop. 4.1, we can prove the correctness of the parallel algorithms in our framework. If the conditions of Prop. 4.1 are not fulfilled, one may have an incorrect parallel FE solver if one does not add additional iterative communication steps [23]. This is illustrated in Fig. 3. Even more than that, $k = \max(1, D(\mathcal{V}_h))$ is the largest possible value for which the parallel FE solver is still guaranteed to be correct, i.e., the value of $k$ that minimizes $|\mathcal{T}|$. As assessed in Sect. 5, and depending on the particular refinement pattern at hand, choosing the largest possible $k$ can lead to a noticeably reduction in the number of mesh cells and computational times (compared to the most conservative value $k = 0$).

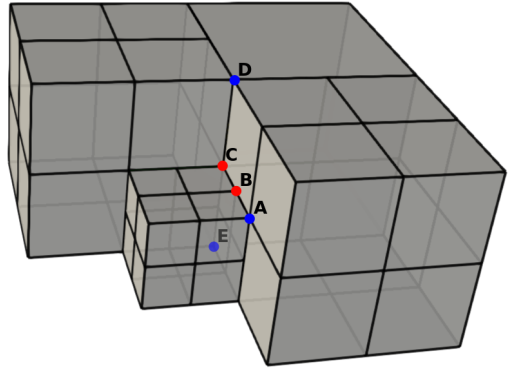

FIGURE 3. 2-balanced forest-of-octrees with 4 octrees (i.e., $|C| = 4$) and a global conforming FE space grounded on trilinear FEs, i.e., $k = 2$ and $D(\mathcal{V}_h) = 0$. For visualization purposes, some of the cells of the mesh are hidden. All these cells are such that $\ell(K) = 1$. In this particular scenario, the conditions of Prop. 4.1 are not fulfilled, as $\max(1, D(\mathcal{V}_h)) = 1 \not\geq k = 2$. The regular and hanging DOFs of interest are marked with blue and red circles, resp. The hanging DOF labeled as $B$ is constrained by $A$ and $C$, which is in turn constrained by $D$ and $E$. Assuming a one-to-one mapping among cells and processors, the processor that owns the cell that contains the edge connecting $A$ and $B$ cannot locally compute the linear constraint associated with $B$, because the cell that contains the edge connecting $D$ and $E$ is not a ghost cell in this processor. We note that in this case the parallel code does not actually crash, but builds a FE space $\mathcal{V}_h$ which is non-conforming, thus, leading to an incorrect parallel FE solver. In order to solve this problem, the root of the octree located in the upper, right corner must be refined so that the forest becomes 1-balanced, and the conditions of Prop. 4.1 are fulfilled.

4.4. **Subdomain-wise assembly of proc-regular interface DOF values.** In a distributed-memory context, non-overlapping domain decomposition solvers need, in the iterative solution process of $\mathbf{Au} = \mathbf{b}$, to assemble, subdomain-wise, the (subassembled) values corresponding to proc-regular interface DOFs (i.e., DOFs in the $\mathcal{N}_R^p \cap \mathcal{N}_\Gamma^p$ set).[8]

---

[8]We note that hanging DOF constraints are eliminated from the system during FE assembly, i.e., we do not allocate an equation/unknown for DOFs in the $\mathcal{N}_H^p \cap \mathcal{N}_L^p$ set.

This operation can be stated as follows. Given a distributed *subassembled* (i.e., partially-summed) vector $\mathbf{x}$, i.e., a vector such that entries of $\mathbf{x}^p$ corresponding to proc-regular interface DOFs hold partial contributions to the corresponding entries in $\mathbf{x}$, *transform $\mathbf{x}$ such that it becomes fully-assembled*, i.e., entries of $\mathbf{x}^p$ corresponding to proc-regular interface DOFs contain the (i.e., fully summed) value of the corresponding entries in $\mathbf{x}$.

We define the set $\mathcal{S}^{\mathrm{proc}}(g) \doteq \{q : g \in \mathcal{N}_R^q \cap \mathcal{N}_\Gamma^q\}$ for any $g \in \mathcal{N}_R^p \cap \mathcal{N}_\Gamma^p$. For each of these DOFs, we assign a *processor owner* among the set of processors in $\mathcal{S}^{\mathrm{proc}}(g)$ (using an algorithm presented later in this section). The owner processor is denoted by $\mathcal{O}^{\mathrm{proc}}(g) \in \mathcal{S}^{\mathrm{proc}}(g)$. The rest of processors in $\mathcal{S}^{\mathrm{proc}}(g)$ become non-owner processors. As required for the exposition, we also define equivalently $\mathcal{S}^{\mathrm{proc}}(F)$ and $\mathcal{O}^{\mathrm{proc}}(F)$ for any VEF $F \in \mathcal{F}_\Gamma^p \cap \mathcal{F}_R^p$. Thus, we can write $\mathcal{S}^{\mathrm{proc}}(g) \doteq \mathcal{S}^{\mathrm{proc}}(F)$ (resp. $\mathcal{O}^{\mathrm{proc}}(g) \doteq \mathcal{O}^{\mathrm{proc}}(F)$), for any $g \in \mathcal{N}_F^p$. In any case, the algorithms in this section do not compute $\mathcal{S}^{\mathrm{proc}}(F)$ and $\mathcal{O}^{\mathrm{proc}}(F)$, but $\mathcal{O}^{\mathrm{proc}}(g)$ and $\mathcal{S}^{\mathrm{proc}}(g)$ directly. This may lead to computational savings, as not all VEFs necessarily own DOFs. (This obviously depends on the FE at hand.) The subdomain-wise assembly operation is split into two communication stages. In the first stage, referred to as **S1**, processor owners receive the corresponding partially-summed contributions from non-owner processors, and reduce-sum the partial sums received into fully-assembled values (at processor owners). In the second stage, referred to as **S2**, processor owners send fully-summed values resulting from the first stage to non-owner processors, so that all processors end up with fully-assembled proc-regular interface DOF values. In the sequel, we sketch the process that sets up the communication patterns underlying **S1** and **S2**.

Each processor needs to figure out the following information in order to set up the **S1** communication pattern. *On the receive side*, $p$ has to determine the set of processor identifiers from which it receives data. Let us denote this set by $\mathcal{S}_{\mathrm{rcv}}^p$. Besides, $p$ needs to figure out, for each processor $q \in \mathcal{S}_{\mathrm{rcv}}^p$, the DOFs whose values have to be accumulated to the (partially-summed) values received from $q$. Let us denote this set by $\mathcal{N}_{\mathrm{rcv}}^{p \leftarrow q}$, for all $q \in \mathcal{S}_{\mathrm{rcv}}^p$. Thus, $|\mathcal{N}_{\mathrm{rcv}}^{p \leftarrow q}|$ is the amount of data items that $p$ receives from $q$. *On the send side*, each processor $p$ has to determine $\mathcal{S}_{\mathrm{snd}}^p$ and $\mathcal{N}_{\mathrm{snd}}^{p \to q}$, for each $q \in \mathcal{S}_{\mathrm{snd}}^p$, i.e., the set of processor identifiers to which $p$ sends data, and the DOFs whose values have to be sent to each processor $q \in \mathcal{S}_{\mathrm{snd}}^p$, resp. Likewise, $|\mathcal{N}_{\mathrm{snd}}^{p \to q}|$ is the amount of vector entries that $p$ sends to $q$. On the other hand, the communication pattern for **S2** can be very easily determined from the one of **S1** by swapping send-side sets and receive-side sets.

Alg. 2 sketches the process in charge of generating the **S1** communication pattern. The inputs of the algorithm are $\mathcal{O}^{\mathrm{proc}}(g)$, for $g \in \mathcal{N}_R^p \cap \mathcal{N}_\Gamma^p$, and $\mathcal{S}^{\mathrm{proc}}(g)$ for $g \in \mathcal{N}_R^p \cap \mathcal{N}_\Gamma^p$ such that $\mathcal{O}^{\mathrm{proc}}(g) = p$. (Note that the algorithm does not actually require $\mathcal{S}^{\mathrm{proc}}(g)$ if $\mathcal{O}^{\mathrm{proc}}(g) \neq p$.) From these inputs, the algorithm generates the $\mathcal{S}_{\mathrm{rcv}}^p$, $\mathcal{N}_{\mathrm{rcv}}^{p \leftarrow q}$, $\mathcal{S}_{\mathrm{snd}}^p$, and $\mathcal{N}_{\mathrm{snd}}^{p \to q}$ sets.

---

**Algorithm 2:** Determine $\mathcal{S}_{\mathrm{rcv}}^p$, $\mathcal{N}_{\mathrm{rcv}}^{p \leftarrow q}$, $\mathcal{S}_{\mathrm{snd}}^p$, and $\mathcal{N}_{\mathrm{snd}}^{p \to q}$.

```
1  for F ∈ F_Γ^p do                                          /* loop over interface VEFs */
2      if F ∈ F_R^p then                                     /* current VEF is proc-regular */
3          for g ∈ N_F^p do
4              q ← O^proc(g)
5              if p = q then                                 /* i am the owner of g */
6                  for q ∈ S^proc(g) \ {p} do
7                      S_rcv^p ← S_rcv^p ∪ {q}; N_rcv^{p←q} ← N_rcv^{p←q} ∪ {g}
8                  end
9              else                                          /* a remote processor q is the owner of g */
10                 S_snd^p ← S_snd^p ∪ {q}; N_snd^{p→q} ← N_snd^{p→q} ∪ {g}
11             end
12         end
13     end
14 end
```

---

It remains to discuss how $\mathcal{O}^{\mathrm{proc}}(g)$ and $\mathcal{S}^{\mathrm{proc}}(g)$ are generated. Alg. 3 presents a sketch of the process in charge of determining $\mathcal{O}^{\mathrm{proc}}(g)$, for $g \in \mathcal{N}_R^p \cap \mathcal{N}_\Gamma^p$, and $\mathcal{S}^{\mathrm{proc}}(g)$ for $g \in \mathcal{N}_R^p \cap \mathcal{N}_\Gamma^p$ such that $\mathcal{O}^{\mathrm{proc}}(g) = p$ (i.e., the input of Alg. 2). Alg. 3 is split into two stages. On the one hand, a local stage spanning Lines 1-13, in which each processor $p$ builds as much as it can of $\mathcal{O}^{\mathrm{proc}}(\cdot)$ and $\mathcal{S}^{\mathrm{proc}}(\cdot)$ only using local information. The local stage of Alg. 3 also walks through proc-hanging interface VEFs; see Line 1 and Lines 8-12. This is justified by the fact that, due to the hanging DOFs constraints, all $g \in \mathcal{N}_{O_F}^p$ become local in all processors that own cells in $\mathcal{T}_F^p$. On the other hand, Alg. 3 features a communication stage spanning Lines 14-30, in which the processors complete the partially constructed version of $\mathcal{O}^{\text{proc}}(\cdot)$ and $\mathcal{S}^{\text{proc}}(\cdot)$ during the first stage. The communication stage is composed by pack (Lines 15-20), nearest-neighbor exchange (Line 21), and unpack (Lines 22-30) steps. In the rest of the section we mathematically prove why this algorithm is correct.

---

**Algorithm 3:** Determine $\mathcal{O}^{\text{proc}}(\cdot)$ and $\mathcal{S}^{\text{proc}}(\cdot)$ for proc-regular interface DOFs.

```
 1  for F ∈ F_Γ^p do                                                    /* loop over interface VEFs */
 2      W ← ⋃_{K ∈ T_F^p} O^proc(K)
 3      if F ∈ F_R^p then                                                /* current VEF is proc-regular */
 4          for g ∈ N_F^p do
 5              S^proc(g) ← S^proc(g) ∪ W
 6              if T_F^p ∩ T_L^p ≠ ∅ then O^proc(g) ← max_{K ∈ T_F^p} O^proc(K)   /* at least one cell surrounding F is local */
 7          end
 8      else                                                              /* current VEF is proc-hanging */
 9          if N_F^p ≠ ∅ then
10              for g ∈ N_{O_F}^p do  S^proc(g) ← S^proc(g) ∪ W           /* loop over DOFs owned by O̅_F */
11          end
12      end
13  end
14  Allocate communication buffers O_buf(·,·) and S_buf(·,·) to size |T^p| × |M_K|
15  for F ∈ F_R^p ∩ F_Γ^p do                                              /* pack local data into communication buffers (Lines 15-20) */
16      for K ∈ T_F^p such that K ∈ T_L^p do                              /* loop over local cells around F */
17          Find f ∈ F_K such that [f] = F
18          for α ∈ M_K^f do  X_buf(K,α) ← X^proc([α]_p), with X = O, S
19      end
20  end
21  Fetch ghost cell data of O_buf(·,·) and S_buf(·,·) from remote processors via nearest-neighbors exchange
22  for K ∈ T_G^p do                                                      /* unpack data from communication buffers (Lines 22-30) */
23      for f ∈ F_K such that [F] ∈ F_L^p ∩ F_R^p do
24          if T_F^p ∩ T_L^p ≠ ∅ then
25              for α ∈ M_K^f do  S^proc([α]_p) ← S^proc([α]_p) ∪ S_buf^p(K,α)   /* augment S^proc(·) with neighbors data */
26          else
27              for α ∈ M_K^f do  O^proc([α]_p) ← O_buf^p(K,α)             /* remote neighbors determine O^proc(·) */
28          end
29      end
30  end
```

---

Let us define the owner processor of a DOF $g \in \mathcal{N}_F^p$ for $F \in \mathcal{F}_\Gamma^p \cap \mathcal{F}_R^p$ as $\mathcal{O}^{\text{proc}}(g) \doteq \max_{K \in \mathcal{T}_F} \mathcal{O}^{\text{proc}}(K)$, with $\mathcal{O}^{\text{proc}}(K)$ denoting the processor owner of $K \in \mathcal{T}$; see Sect. 3.2. The following result holds.

**Proposition 4.2.** *Let us consider a distributed forest-of-trees mesh $\mathcal{T}$ with s-ghost cells, where $D(\mathcal{V}_h) \geq s$. For any processor $p$, the owner processor of $g \in \mathcal{N}_F^p$, for $F \in \mathcal{F}_\Gamma^p \cap \mathcal{F}_R^p$, can be locally computed as $\mathcal{O}^{\text{proc}}(g) = \max_{K \in \mathcal{T}_F^p} \mathcal{O}^{\text{proc}}(K)$ if $\mathcal{T}_F^p \cap \mathcal{T}_L^p \neq \emptyset$ (see Line 6). Otherwise (i.e., $\mathcal{T}_F^p \cap \mathcal{T}_L^p = \emptyset$), any neighbor processor that owns a ghost cell in $\mathcal{T}_F^p$ can compute $\mathcal{O}^{\text{proc}}(g)$ (and, thus, $p$ can fetch this information from any of these neighbors).*

As a result of this proposition, the communication stage in Alg. 3 (Lines 14-30) is guaranteed to obtain $\mathcal{O}^{\text{proc}}(g)$ on processors such that $\mathcal{T}_F^p \cap \mathcal{T}_L^p = \emptyset$ as well.

On the other hand, with regard to $\mathcal{S}^{\text{proc}}(g)$, we note that a processor that is owner of a proc-regular interface DOF $g$, i.e., $\mathcal{O}^{\text{proc}}(g) = p$, may not able to determine the full $\mathcal{S}^{\text{proc}}(g)$ set solely using local information (i.e., local cells plus a layer of $s$-ghost cells). This scenario is illustrated in Fig. 4c for the DOF at processor $p = 5$ pointed by an arrow. Using only local information, $\mathcal{S}_5(g) = \{3, 4, 5\}$. However, processor 2 is an element of $\mathcal{S}_5(g)$ as well. The cell owned by processor 2 that would let processor 5 complete the $\mathcal{S}_5(g)$ set is not in the ghost cells set of processor 5. However, processor 5 retrieves such missing local information fetching data from processor 3 in the processor 5 ghost cells which are owned by processor 3, in particular in Line 25 of Alg. 3. The approach followed by Alg. 3 in order to build $\mathcal{S}^{\text{proc}}(g)$ is correct as proved in the following proposition.

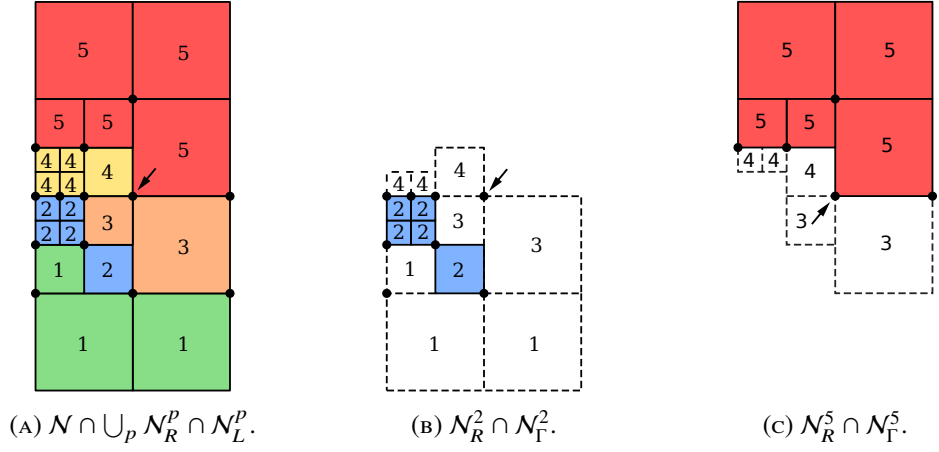

(A) $\mathcal{N} \cap \bigcup_P \mathcal{N}_R^p \cap \mathcal{N}_L^p$.   (B) $\mathcal{N}_R^2 \cap \mathcal{N}_\Gamma^2$.   (C) $\mathcal{N}_R^5 \cap \mathcal{N}_\Gamma^5$.

FIGURE 4. Proc-regular interface DOFs (black circles) of a global conforming FE space grounded on bilinear Lagrangian FEs on top of a 2:1 0-balanced forest-of-quadtrees mesh with two quadtrees distributed (non-uniformly) among $P = 5$ processors with 0-ghost cells. The DOF pointed out by an arrow in Fig. 4c is such that processor $p = 5$, even being its owner, is not able to determine the full $\mathcal{S}^{\text{proc}}(g)$ set solely using local information. Fortunately, as stated by Prop. 4.3, this set can be obtained combining local information and a single nearest neighbor communication.

**Proposition 4.3.** *Let us consider a conforming FE space $\mathcal{V}_h$ and a distributed 2:1 k-balanced forest-of-trees mesh $\mathcal{T}$ with k-ghost cells, where $\max(1, D(\mathcal{V}_h)) \geq k$. Then, for any $g \in \mathcal{N}_R^p \cap \mathcal{N}_\Gamma^p$ such that $O^{\text{proc}}(g) = p$, the full $\mathcal{S}^{\text{proc}}(g)$ set can be computed combining local information and a single nearest-neighbor communication.*

**Remark 4.4** (Implementation remark). *In order to keep the presentation short, we have omitted from Alg. 2 and 3 that DOFs in $\mathcal{N}_{\text{snd}}^{p \to q}$ and $\mathcal{N}_{\text{rcv}}^{q \leftarrow p}$, for any pair of neighbors p and q, have to be glued together at both sides of the interface, as a final step to complete the equivalence classes corresponding to DOFs at the subdomain interfaces. We achieve this is in practice as follows. Alg. 3 generates, in situ, a global DOF identifier for all DOFs $g \in \mathcal{N}_R^p \cap \mathcal{N}_\Gamma^p$. The algorithm that generates such identifiers follows the same pattern to the one that determines $O^{\text{proc}}(g)$, and thus is mathematically supported by Prop. 4.2 as well. In particular, if a proc-regular interface DOF is surrounded by at least one local cell, a global DOF identifier can be generated solely using processor-local information, while if not, it can be fetched on ghost cells via a nearest-neighbor exchange. Then, using such global DOF identifiers, Alg. 2 arranges the DOFs in $\mathcal{N}_{\text{snd}}^{p \to q}$ and $\mathcal{N}_{\text{rcv}}^{q \leftarrow p}$ in increasing global DOF numbering within each set, such that they are consistently paired at both sides of the interface among p and q. The global DOF identifiers are then discarded, and not used elsewhere.*

4.5. **Setting up parallel distributed fully-assembled linear systems.** The vast majority of parallel linear algebra packages (e.g., PETSc [18], or TRILINOS [19]) use a data distribution model in which the global linear system is partitioned by rows, i.e., each processor *owns* a non-overlapping subset of *fully-assembled* global rows of $\mathbf{A}$ and $\mathbf{b}$, resp. Besides, the data structures are *globally addressable*, i.e., a given processor may contribute to a global matrix row (or a global vector entry) which is not necessarily owned by it. Such packages can be leveraged by means of a distributed algorithm able to generate the equivalence classes in $\mathcal{N}_R$ with the local-to-global index map [·] built such that the DOFs owned by the first processor are the ones to be numbered first, immediately followed by those of the second, and so on.[9]

The distributed algorithm presented in [14, Sect. 3.1] (available in deal.II) follows a cell-wise approach that exploits the overlapped mesh partition to directly generate such index map. Here, we instead follow a DOF-wise approach which exploits the locally generated local-to-proc-local index map $[·]_p$ (see Sect. 4.2) and Remark 4.4 to assign a global DOF identifier to DOFs in $\mathcal{N}_L^p \cap \mathcal{N}_R^p$, for all $p$. We define $\mathcal{N}_O^p \doteq \{g \in \mathcal{N}_L^p \cap \mathcal{N}_R^p : O^{\text{proc}}(g) = p\}$, and refer to it as the set of DOFs owned by processor $p$. This algorithm encompasses four main steps: (a) each processor locally computes $|\mathcal{N}_O^p|$; (b) an exclusive prefix reduction (i.e., MPI_Exscan) of $|\mathcal{N}_O^p|$, for all $p$, computes a starting offset global DOF identifier on each processor; (c) each processor locally assigns a global, consecutive DOF identifier to owned DOFs starting from the global offset identifier computed in (b); (d) each processor fetches from remote neighbors the global DOF identifiers for those DOFs in $\mathcal{N}_L^p \cap \mathcal{N}_R^p$ which are not owned by it. This operation involves a DOF-wise nearest neighbor communication with $\mathcal{S}_{\text{rcv}}^{p}$, $\mathcal{N}_{\text{rcv}}^{p \leftarrow q}$ as send side, and $\mathcal{S}_{\text{snd}}^{p}$, $\mathcal{N}_{\text{snd}}^{p \rightarrow q}$ as receive side (i.e., owners send, non-owners receive).

---

[9]We note that to have consecutive global row identifiers in each processor is a constraint of PETSc [18], but not actually TRILINOS [19].

## 5. Numerical experiments

5.1. **Experimental environment.** The numerical experiments are run at the Marenostrum-IV (MN-IV) supercomputer, hosted by BSC. This petascale machine is equipped with 3,456 compute nodes interconnected together with the Intel OPA HPC network. Each node has 2x Intel Xeon Platinum 8160 multi-core CPUs (Skylake), with 24 cores each (i.e. 48 cores per node) and 96 GBytes of RAM.

With respect to the software, we used `FEMPAR` v1.0.0 [20], linked against `p4est` v2.2 [2, 3] as its forest-of-octrees manipulation engine. In its current status, up to the authors' knowledge, `t8code` [1] only provides facet-oriented variants of the main operations outlined in Sect. 2.6 (e.g., `Balance` is restricted to $d - 1$-balance), and thus cannot be readily used for the implementation of generic FEs. For this reason, in this paper we restrict to `p4est` as a practical demonstrator. Besides, we used `deal.II` v9.0.0 [15], linked against `p4est` v2.2, and `PETSc` v3.9.0 [18] for distributed-memory linear algebra data structures and solvers. These software were compiled with Intel v18.0.1 compilers using system recommended optimization flags and linked against the Intel Message Passing Interface (MPI) Library (v2018.1.163) for message-passing and the BLAS/LAPACK and PARDISO available on the Intel MKL library for optimized dense linear algebra kernels and sparse direct solvers, resp. All floating-point operations were performed in IEEE double precision.

5.2. **Poisson problem.** We evaluate the performance and strong scalability of the different stages in our framework as implemented in `FEMPAR`, and compare them against those in `deal.II` for the numerical solution of the Poisson problem. We consider a 3D cube domain $\Omega = [0, 1]^3$ with homogeneous Dirichlet boundary conditions (strongly) imposed over the entire domain boundary. The problem is discretized using a global Lagrangian FE space $\mathcal{V}_h \subset H_0^1(\Omega)$. The weak formulation of this problem reads: find $u_h \in \mathcal{V}_h$ such that

$$(\nabla u_h, \nabla v_h) = (f, v_h), \qquad \forall v_h \in \mathcal{V}_h. \tag{3}$$

The right-hand-size $f(x, y, z)$ is a piece-wise function defined as:

$$f(x, y, z) = \begin{cases} 1 & z > \frac{1}{2} + \frac{1}{4} \sin(4\pi x) \sin(4\pi y) \\ -1 & z \leq \frac{1}{2} + \frac{1}{4} \sin(4\pi x) \sin(4\pi y) \end{cases}.$$

The discontinuity in $f$ leads to a solution that is non-smooth along a sinusoidal surface through the domain, so that very localized AMR is required in order to reduce the error in that area, while keeping the computational requirements reasonably low.

The experiment is designed as follows. We start with a coarse mesh that is obtained after uniformly refining the root octant of the octree up to four times. This mesh has 16 cells (hexahedra) per coordinate direction, and 4096 cells in total. Then, the Poisson problem is solved on a hierarchy of meshes where the mesh at a given step is obtained from the previous one by means of a user-level mesh handling primitive in our framework named `Refine_and_coarsen`. This primitive adapts the forest-of-trees mesh based on cell flags set by the user, while transferring data that the user might have attached to the mesh objects (i.e., cells and VEFs) to the new mesh objects generated after mesh adaptation. `Refine_and_coarsen` invokes `Adapt`, `Balance`, and `Ghost`; see Sect. 2.6. With load balancing in mind, the mesh is dynamically redistributed at each step by means of the `Redistribute` mesh handling primitive right after each call to `Refine_and_coarsen`. This primitive dynamically balances the number of cells in each processor. The data that the user might have attached to the mesh objects (i.e., cells and VEFs) is also migrated among processors. `Redistribute` invokes `Partition`, and `Ghost`; see Sect. 2.6. We note that both `Refine_and_coarsen` and `Redistribute` also have to: (a) set up the data structures providing cell neighbors across $n$-faces; (b) call Alg. 1. The process underlying (a) is implemented by means of invocations to `Iterate` (Sect. 2.6).

For a given (fixed) number of AMR steps, we measure the elapsed time (i.e., wall clock time) spent in each of the stages in the simulation of process, *aggregated across all steps*, and evaluate at which rate they are reduced with increasing number of CPU cores (i.e., strong scalability test). The number of AMR steps is adjusted such that a "sufficiently large" problem size at the last step is obtained for the CPU core range on which we run the strong scaling test. The decision of which cells to be refined or coarsened is performed as follows. Given the solution of the linear system, each processor computes independently of each other an error indicator for cells $K \in \mathcal{T}_L^p$ using the a-posteriori error estimator proposed by Kelly et al. in [24]. Then, given user-defined refinement and coarsening fractions, denoted by $\alpha_r$ and $\alpha_c$, resp., we find the thresholds $\theta_r$ and $\theta_c$ such that the *total* number of cells with error indicator larger (resp., smaller) than $\theta_r$ (resp., $\theta_c$) is (approximately) a fraction $\alpha_r$ (resp., $\alpha_c$) of the *total* number of cells. To this end, we use the iterative algorithm proposed in [14, Fig. 5]. We set $\alpha_r = 0.15$ and $\alpha_c = 0.03$, so that, the number of cells is at least doubled at each AMR step (assuming that no cells can be coarsened). The actual number of cells at each mesh in the hierarchy, however, also depends on the algorithm within p4est that 2:1 balances the forest-of-octrees, as it may need to apply more refinement in order to keep this constraint [2].

In the sequel, we will focus on evaluating the following stages, which are labeled as:

- MESH: Includes the calls to the Refine_and_coarsen and Partition primitives.
- FE SPACE SUB-ASSEMBLY: Includes the generation of proc-local DOFs (Sect. 4.2), Alg. 3, and the computation of hanging DOFs constraints for all proc-local DOFs $g \in \mathcal{N}_H^p \cap \mathcal{N}_L^p$.
- FE SPACE FULL-ASSEMBLY: Includes FE SPACE SUB-ASSEMBLY and the algorithm outlined in Sect. 4.5.
- ASSEMBLY SUB-ASSEMBLY: Includes the computation of the cell matrix and force for all local cells, the treatment of hanging DOF constraints according to [13], and the incremental assembly and final compression of the local-to-subdomain coefficient matrix data structure; see [20, Sect. 11.1] for more details on this last step.
- ASSEMBLY FULL-ASSEMBLY: Includes ASSEMBLY SUB-ASSEMBLY and the setup of the fully-assembled linear system (using the data structures in PETSc). This latter step is split into three substeps: (A) the sparsity pattern of locally owned rows is completed by means of a nearest neighbor exchange; (B) the full sparsity pattern of locally owned rows computed in (A) is used to set up the fully-assembled matrix data structure; (C) the subassembled non-zero matrix entries and vector entries are injected, on a row-by-row fashion, into the global, fully-assembled data structures. Then, a final inter-processor assembly stage involves the transfer of partial contributions to local entries but actually stored in (i.e., owned by) remote processors. This communication phase is resolved by the external linear algebra package.
- ERROR ESTIMATOR: Includes the computation of the error estimators for all local cells, the computation of $\theta_r$ and $\theta_c$ thresholds, and to flag the cells for the Refine_and_coarsen primitive.
- TOTAL SUB-ASSEMBLY: The aggregation of MESH, FE SPACE SUB-ASSEMBLY, ASSEMBLY SUB-ASSEMBLY, and ERROR ESTIMATOR.
- TOTAL FULL-ASSEMBLY: The aggregation of MESH, FE SPACE FULL-ASSEMBLY, ASSEMBLY FULL-ASSEMBLY, and ERROR ESTIMATOR.

In this section, we also compare the performance and scalability of the algorithms and data structures within FEMPAR with those in deal.II. In particular, we adapted the deal.II step-40 tutorial program [10] in order to solve problem (3), and grouped the different steps in this program into stages according to the classification above. There are, though, still several significant worth-noting differences among the following stages:

(1) The MESH stage in deal.II does not actually call Alg. 1, but a fundamentally different algorithm; see Remark 3.13.
(2) The FE SPACE FULL-ASSEMBLY stage in deal.II does not use the steps in FE SPACE SUB-ASSEMBLY, but the algorithm in [14, Sect. 3.1]. Besides, it also sets up the so-called *locally owned* and *locally relevant* index sets of global DOF identifiers. These data structures are required to manipulate globally addressable, fully-assembled linear algebra matrices and vectors in deal.II. We refer to [14, Sect. 3] for additional details.
(3) The ASSEMBLY FULL-ASSEMBLY stage in deal.II, as its counterpart in FEMPAR, comprises three different steps, with step (B) in deal.II being equivalent to its FEMPAR counterpart. Step (A) first determines, *by means of symbolic assembly*, the location of non-zero elements in the local subassembled matrix and then, a communication stage, equivalent to the one in FEMPAR above, allows all processors to complete the non-zero pattern of locally owned rows; in step (C) the contribution of local cells are computed and *directly* assembled into the global linear system data structures, including the transfer of entries locally computed but stored in remote processors. We refer to [14, Sect. 4] for additional details.

We do not report the computational time spent in the linear system solve stage, but focus on the algorithms presented in this paper instead.

---

[10]Documentation available at https://www.dealii.org/9.0.0/doxygen/deal.II/step_40.html.

We note that, for the problem at hand, $D(\mathcal{V}_h) = 0$. Thus, due to Prop. 4.1, and 4.2, we may use $k = 1$ and $s = 0$. However, we use $k = 0$ and $s = 0$ to have a fair comparison against `deal.II`.[11] The experiments with $k = 1$ and $s = 0$ reveal a *negligible* reduction in the number of adaptive mesh cells in the case of the experiments reported in Fig. 5 (less than 0.02% for the mesh in the last AMR step), but *a much more noticeable one* for the ones in Fig. 6 (up to 12.1% for the mesh in the last AMR step).

Fig. 5 (left) shows the strong scalability of the algorithms and data structures in `FEMPAR` for problem (3) discretized with trilinear ($Q_1(K)$) Lagrangian FEs. On the other hand, Fig. 5 (right) shows the ratio, $R$, among the computational times spent in the stages in `deal.II` and their counterparts in `FEMPAR`. If $R$ is larger than one, then the corresponding stage in `FEMPAR` is faster than the one in `deal.II` by a factor $R$. For readability purposes, Fig. 5 (left) also provides the ideal strong scaling slope (solid black line). The more parallel a given strong scaling curve is to the ideal slope the more strongly scalable the corresponding stage is. Two different global problem sizes suitable for the [48, 12228] and [384, 30672] CPU cores range were tested. In particular, the results in Fig. 5a correspond to a problem in which we perform 13 AMR steps, resulting into a 49.8 MDOFs problem size at the last step, while in Fig. 5b, we report the ones for 16 AMR steps, and a 415.5 MDOFs problem size.[12] *These DOF counts, and the ones in the rest of the paper, include both regular and hanging DOFs.* The average load per core at the last step ranges, in Fig. 5a, from 1.04M DOFs/core to 4.06K DOFs/core, and, in Fig. 5b, from 1.08M DOFs/core to 13.53K DOFs/core. For readability purposes, we also plot a vertical line corresponding to the 25K DOFs/core regime. Beyond that point, one can barely expect AMG linear solvers to scale.

Fig. 5a and 5b (left) reveal high strong scalability for the ASSEMBLY SUB-ASSEMBLY and ERROR ESTIMATOR stages in the full CPU cores range tested. For example, the parallel efficiency for the ASSEMBLY SUB-ASSEMBLY stage only decays to 74% and 78% with the largest number of cores tested in Fig. 5a and Fig. 5b, resp. An intermediate degree of strong scalability is observed for the ASSEMBLY FULL-ASSEMBLY stage, that decays to 50% and 44% in Fig. 5a and Fig. 5b, resp., due to the communication overhead involved in stages (A) and (C) outlined above. Also worth noting is the extra overhead of ASSEMBLY FULL-ASSEMBLY compared to ASSEMBLY SUB-ASSEMBLY, which ranges from 25% for 48 cores to 85% for 12.2K cores in Fig. 5a, and from 15% for 384 to 100% for 30.7K cores in Fig. 5a. This is caused by the extra stages (A)-(C) involved in the setup of fully-assembled linear systems. While less strong scalability is observed for the MESH, FE SPACE SUB-assembly, and FE SPACE FULL-assembly stages, the computational time reduction with the number of cores is still very significant for these stages, with the parallel efficiency decaying to 16%, 21%, and 19%, in Fig. 5a, and 26%, 36%, and 23%, resp., in Fig. 5b. This is not surprising as these three stages, and specially MESH, involves significantly more communication volume (relative to local work) than, e.g., ASSEMBLY SUB-ASSEMBLY.

In Fig. 5a and 5b (right), we can observe that the algorithms and data structures in `FEMPAR` are either competitive (ERROR ESTIMATOR and ASSEMBLY FULL-ASSEMBLY stages) or *up to 2-3 times faster* than the ones in `deal.II` (MESH, FE SPACE FULL-assembly). The ratio in the MESH stage is, e.g., as large as 2.08x for the largest number of cores in Fig. 5a, and increases at a moderate pace from 1.70x to 2.08x from 48 to 12.2K cores. These results evidence what it seems a more efficient coupling of the data structures in `FEMPAR` and those of `p4est`, although we could not completely discard whether the performance improvement is caused by differences among low-level technical details in the HPC implementation of both codes and the underlying hardware. Even larger ratios for the FE SPACE FULL-assembly stage are observed in Fig. 5a and 5b, as large as 2.75x, and 2.49x, resp., for the largest number of cores evaluated. The extra communication volume and overhead associated to handling a global numbering of DOFs suitable for fully-assembled operators in `deal.II` (e.g., searches on index sets of global identifiers, 64-bit DOF identifiers) is the most contributing factor to this trend.

If we focus on the ratios for the ASSEMBLY FULL-ASSEMBLY stage in Fig. 5a and 5b, one can observe a sudden parallel scaling drop of `deal.II` compared to `FEMPAR`. We confirmed that the source of this difference is concentrated in steps (A) and (C) of ASSEMBLY FULL-ASSEMBLY in `deal.II`. We could not, however, determine the actual cause of the difference. In any case, we consider this issue to be of relative importance provided that this effect arises at a regime in which one has a relatively low load per core. Indeed, this was actually not observed in [14] (at least as

---

[11]`deal.II` always instructs `p4est` to build 2:1 0-balanced meshes, i.e., $k$ is not a parameter the user can play around with. See line 2873 of https://www.dealii.org/9.0.1/doxygen/deal.II/distributed_2tria_8cc_source.html for more details.

[12]We stress that this is by no means a parallel scaling limit of the algorithms proposed, but the largest number of cores we can exploit on MN-IV provided the access constraints that we have to this supercomputer. We expect the algorithms proposed to be able to scale up to hundreds of thousands of cores in the solution of hundred of billions DOFs problem sizes.

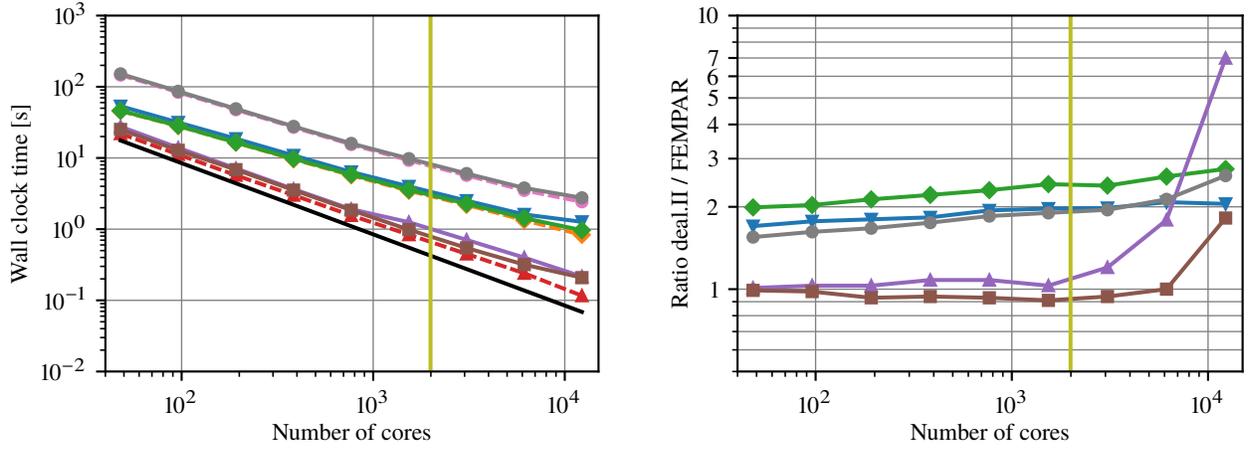

(A) 13 AMR steps, 49.8 MDOFs, up to 12.2K cores.

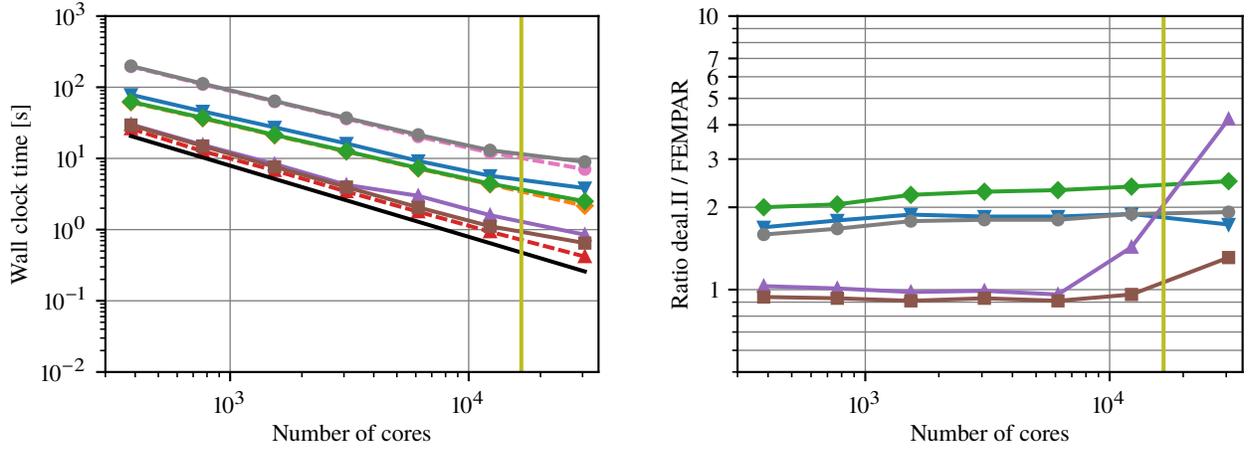

(B) 16 AMR steps, 415.5 MDOFs, up to 30.7K cores.

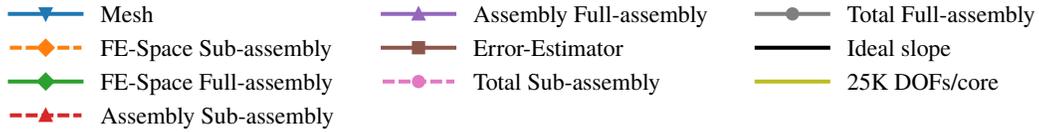

FIGURE 5. Strong scaling test results on MN-IV for `FEMPAR` (left) and ratio `deal.II` versus `FEMPAR` (right). Poisson problem with $Q_1(K)$ Lagrangian FEs.

evidently). This might be caused by the fact that experiments in [14] considered strong scaling results up to larger loads per core to those considered here.

Overall, *the balance achieved among all these stages ends up with* TOTAL FULL-ASSEMBLY *in `FEMPAR` being up to 2.60x faster in Fig. 5a, and 1.92x faster in Fig. 5b compared to* TOTAL FULL-ASSEMBLY *in `deal.II` (for the largest number of cores tested).*

In order to showcase the suitability of the proposed $h$-adaptive pipeline for higher than linear-order FEs, Fig. 6 shows its strong scalability for problem (3) discretized with triquadratic ($Q_2(K)$) Lagrangian FEs. The average load per core at the last step ranges, in Fig. 6a, from 0.92M DOFs/core to 3.57K DOFs/core, and, in Fig. 6b, from 1.25M DOFs/core to 15.72K DOFs/core. These loads per core are similar to those in Fig. 5. However, for a given number of cores, the load per core in terms of number of mesh cells at the last step, is smaller in Fig. 6, compared to the one in Fig. 5. The results in Fig. 6a correspond to a problem in which 9 AMR steps are performed (4 steps less than with trilinear FEs), resulting into a 43.9 M DOFs problem size at the last step, while in Fig. 6b, we have 12 AMR steps (3 steps less), and a 482.2 MDOFs problem size.

All stages in Fig. 6 are less strongly scalable than their counterparts in Fig. 5, except for the ASSEMBLY SUB-ASSEMBLY stage, which is as scalable as in Fig. 5, and, by far, the most strongly scalable stage among those in Fig. 6.

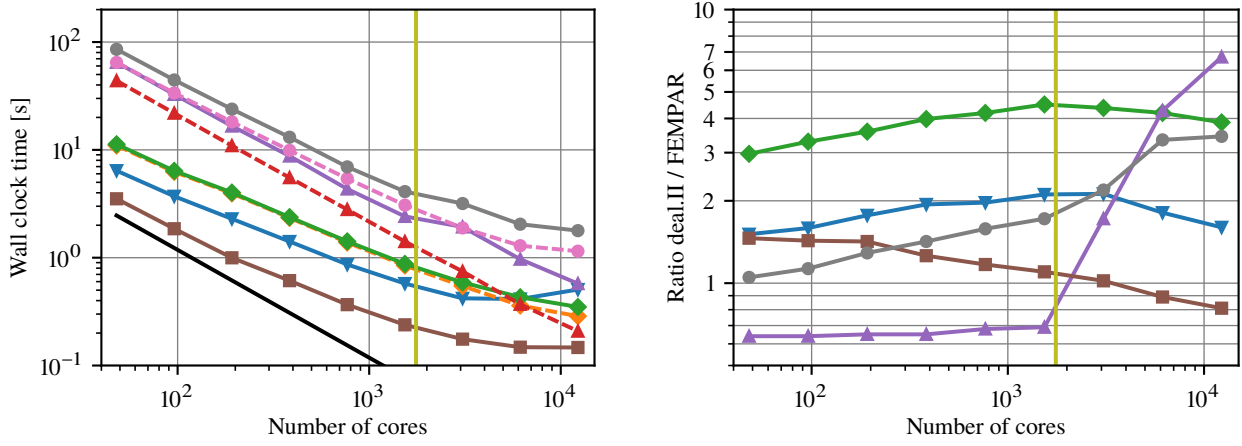

(A) 9 AMR steps, $Q_2(K)$ FEs, 43.9 MDOFs, up to 12.2K cores.

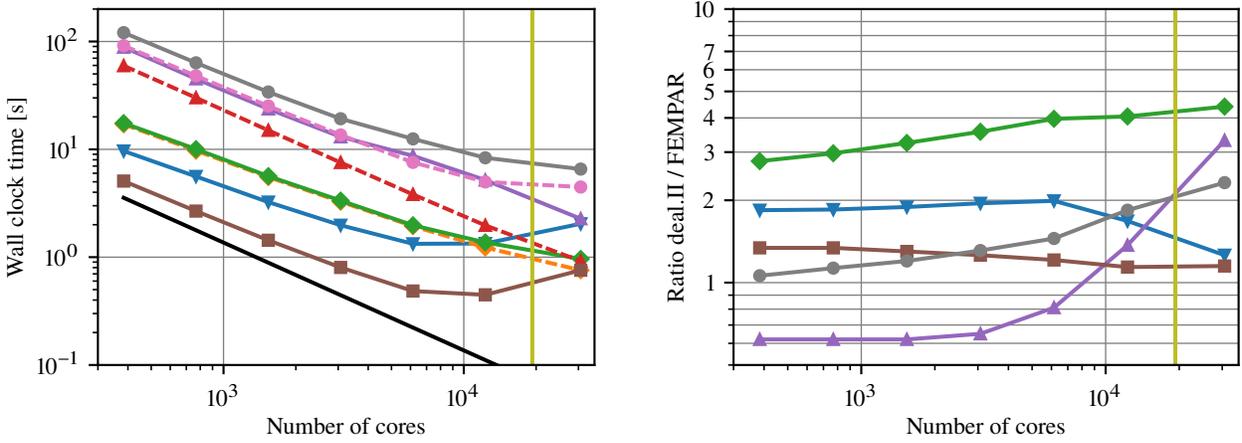

(B) 12 AMR steps, $Q_2(K)$ FEs, 482.2 MDOFs, up to 30.7K cores.

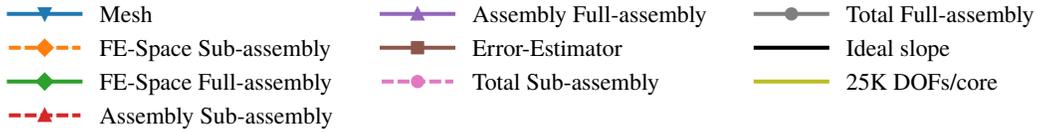

FIGURE 6. Strong scaling test results on MN-IV for `FEMPAR` (left) and ratio `deal.II` versus `FEMPAR` (right). Poisson problem with $Q_2(K)$ Lagrangian FEs.

In any case, the computational time reduction with the number of cores is still significant for the rest of stages, except for the MESH and ERROR ESTIMATOR stages and the two largest core counts, for which any benefit of parallelism is totally absorbed by parallel overheads. Also worth noting in Fig. 6 is that ASSEMBLY SUB-ASSEMBLY and ASSEMBLY FULL-ASSEMBLY concentrate a higher percentage of TOTAL SUB-ASSEMBLY and TOTAL FULL-ASSEMBLY, resp. (compared to Fig. 5). This hides the higher drop of parallel efficiency that is observed for the MESH and ERROR ESTIMATOR stages.

If we compare the performance and scalability of `FEMPAR` and `deal.II` stages in Fig. 6a and 6b (right), rather different conclusions can be extracted depending on the stage. The MESH stage in `FEMPAR` is faster than the one in `deal.II` for the whole CPU core range evaluated. However, while it scales better within a first range of CPU cores (e.g., in the [48, 6.1K] range in Fig. 6a), a drop of the ratio `deal.II` versus `FEMPAR` is observed for this stage in Fig. 6a and Fig. 6b, with the drop in the latter being more significant than the one of the former. On the contrary, ASSEMBLY FULL-ASSEMBLY stage in `deal.II` is 1.6x faster than its counterpart in `FEMPAR` within a first range of CPU cores in which ASSEMBLY FULL-ASSEMBLY is dominated by local work; this reveals that more effort should be invested in the optimization of stages (A)-(C) for $Q_2(K)$ FEs, for which, at present, `FEMPAR` only provides a naive, reference implementation. However, provided the number of CPU cores is sufficiently large, the parallel efficiency of the Assembly Full-Assembly stage in `deal.II` starts dropping significantly compared to the one in `FEMPAR`, up to an extent that large `deal.II` versus `FEMPAR` ratios are observed.

The FE Space Full-assembly stage in `FEMPAR` is faster and scales better than the one `deal.II` for the whole CPU core range evaluated. Indeed, larger ratios for the FE Space Full-assembly stage (and higher ratio increase with the number processors) are observed in Fig. 6 compared to those in Fig. 5. For example, the ratio in the FE Space Full-assembly stage is as large as 4.4x for the largest number of cores in Fig. 6b, and increases from 2.78x to 4.4x from 384 to 30.7K cores due to the extra communication volume and overhead associated to handling a global numbering of DOFs suitable for fully-assembled operators in `deal.II`. The Error Estimator stage is faster in `FEMPAR` within the whole CPU cores range in Fig. 6b, and the same applies to Fig. 6a except for the two largest number of cores. However, the drop in parallel efficiency observed in the left part of Fig. 5b and Fig. 6b is accompanied with a drop of the `FEMPAR` to `deal.II` ratio, moderate in the right part of Fig. 6b (in particular, a 14% ratio drop), but much more apparent in the right part of Fig. 5b (a 42% ratio drop). After thorough inspection with the help of performance analysis tools, we could confirm so far that the drop is not caused by the algorithm which determines the $\theta_r$ and $\theta_c$ thresholds, but in the actual numerical computations performed locally within each processor during this stage.

Overall, *the balance achieved among all these stages ends up with* Total Full-assembly *in* `FEMPAR` *being up to 3.4x faster in Fig. 5a and 2.3x faster in Fig. 5b compared to* Total Full-assembly *in* `deal.II` *(for the largest number of cores tested).*

5.3. **Maxwell problem.** In this section we showcase the application of our framework to the numerical solution of the Maxwell equations. We consider a 3D L-shaped domain $\Omega = [-1,1]^3 \setminus [-1,0]^3$ with non-homogeneous Dirichlet boundary conditions (strongly) imposed over the entire domain boundary. The problem is discretized using a global space $\mathcal{V}_h$ that conforms to $H_0(\mathbf{curl}, \Omega) \doteq \{\mathbf{u} \in H(\mathbf{curl}, \Omega) : \mathbf{n} \times \mathbf{u} = \mathbf{0} \text{ on } \partial\Omega_D\}$. Such $\mathcal{V}_h$ is built from local Nédélec (a.k.a. edge) FEs of first kind. (We refer to [22] for a comprehensive coverage on the general software implementation of this sort of FEs within `FEMPAR`.) The problem reads: find $\boldsymbol{u}_h^0 \in \mathcal{V}_h$ such that

$$(\nabla \times \boldsymbol{u}_h^0, \nabla \times \boldsymbol{v}_h) + (\boldsymbol{u}_h^0, \boldsymbol{v}_h) = (\boldsymbol{f}, \boldsymbol{v}_h) - (\nabla \times E u_D, \nabla \times \boldsymbol{v}_h) + (E u_D, \boldsymbol{v}_h), \qquad \forall \boldsymbol{v}_h \in \mathcal{V}_h, \qquad (4)$$

where $\boldsymbol{u}_D$ is the Dirichlet data to be imposed, and $Eu_D$ an arbitrary extension, i.e., $Eu_D = u_D$ on $\partial\Omega_D$. The magnetic solution field is given by $\boldsymbol{u}_h \doteq \boldsymbol{u}_h^0 + Eu_D$. The input data $\boldsymbol{u}_D$ and $\boldsymbol{f}$ are set up such that the (analytical) solution of this problem becomes $\boldsymbol{u} = \nabla \left( r^{\frac{2}{3}} \sin\left(\frac{2t}{3}\right) \right)$, with $t = \arccos\left(\frac{xyz}{r}\right)$, $(x,y,z)$ the Cartesian coordinates of any point within $\Omega$, and $r$ is the corresponding radius in 3D polar coordinates. This analytical solution has a singular behavior near the origin and $\boldsymbol{u} \notin H^1(\Omega)$. We note that, for this problem, a forest-of-octrees with three adaptive octrees is sufficient in order to geometrically discretize $\Omega$. Besides, as it has known analytical solution, we do not actually use a-posteriori error estimators, but the true error among the analytical and FE solution measured in the $L_2$-norm instead. For this sort of problem, up to the authors' knowledge, there is no `deal.II` step tutorial available at the public domain, so that we could not drive the comparison of `FEMPAR` against `deal.II`. In any case, `deal.II` supports edge $h$-adaptive FEs as well.

Fig. 7 (left) shows the strong scalability of `FEMPAR` for problem (4) discretized with first-order Edge FEs. A single global problem size suitable for the [48, 12228] CPU cores range was tested, in particular, the one corresponding to 11 AMR steps, which already leads to a problem with 115.6 MDOFs at the last step. The averaged load per core at the last step ranges from 2.41M DOFs/core (48 processors) to 9.40K DOFs/core (12.2K processors). The computational times in Fig. 7 (left) were obtained with $k = 0$ and $s = 0$. However, we note that, as $D(\mathcal{V}_h) = 1$, we can use, due to Prop. 4.1 and 4.2, $k = 1$ and $s = 1$. A reduction of 4.5% in the number of cells of the mesh in the last AMR step was observed. Besides, Fig. 7 (right) evaluates the ratio of the computing times obtained with $k = 0$ and $s = 0$ versus the ones obtained with $k = 1$ and $s = 1$.

Fig. 7 (left) confirms, as with the Poisson problem discretized with Lagrangian FEs, remarkable strong scalability for all stages, and significant computational time reductions for Total Sub-assembly in the whole range of CPU cores analyzed, despite the more general structure and complexity of implementation underlying $H$-curl conforming FE spaces. On the other hand, the computational savings in Fig. 7 (right), which are enabled by Prop. 4.1 and 4.2, are such that Total Sub-assembly becomes approximately 1.2 times faster with $k = 1$ and $s = 1$.

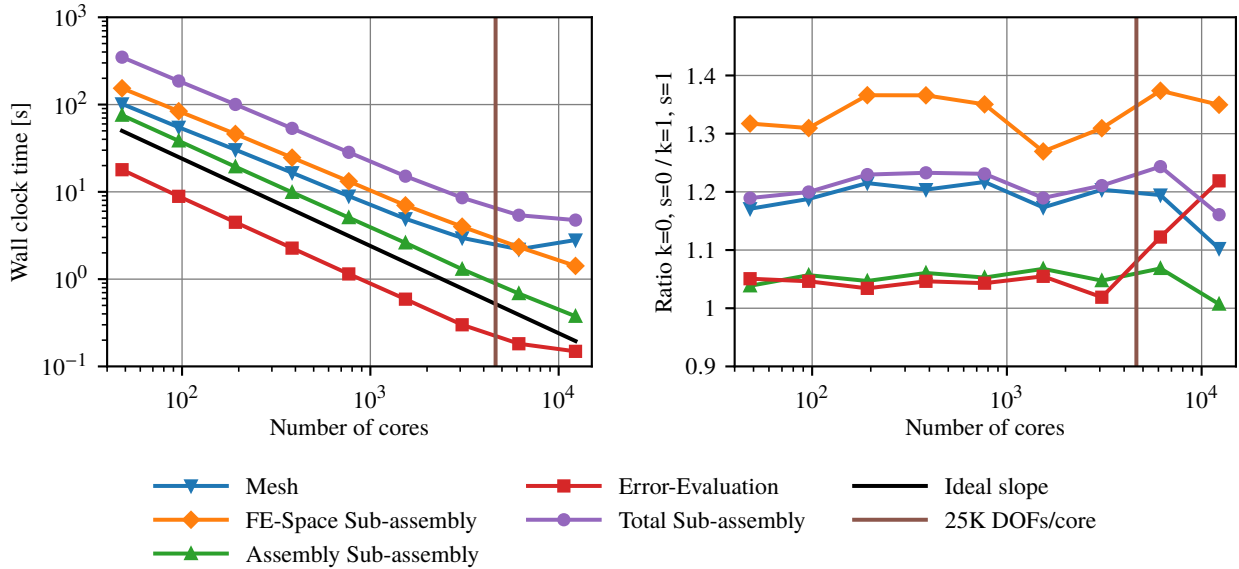

FIGURE 7. FEMPAR strong scaling test results on MN-IV (left) and ratio among computing times with 0-balance and 0-ghost cells versus 1-balance and 1-ghost cells (right). Maxwell problem with first order ($\mathcal{ND}_1(K)$) Edge FEs.

## 6. Conclusions

In this paper we have presented the building blocks, i.e., data structures and algorithms, of a highly scalable parallel *generic FE* simulation framework that supports AMR via forest-of-trees endowed with SFCs. In particular, we have mathematically proven the correctness of the algorithms for scalable mesh handling, construction of generic global conforming FE spaces on top of non-conforming meshes using hanging DOF constraints, and the parallel subdomain-wise sub-assembly and full-assembly of the discrete linear system of algebraic equations. Central to the framework is a generic FE-suitable adaptive mesh representation that is grounded on concepts that are not tailored to a particular cell topology, SFC, or number of space dimensions. All that the framework requires from the forest-of-trees layer meshing engine to be able to build such mesh data structure is a description of the local neighborhood across the cell boundaries in the adapted mesh. Along the way, we have (a) mathematically justified which are the crucial benefits (i.e., ease of implementation and parallelization, high scalability) that one enjoys from enforcing the 2:1 $k$-balance constraint, (b) identified what a conforming FE formulation must fulfill such that it can be implemented in parallel given the constraint that each processor only has access to the off-processor cells in the $s$-ghost cell set, and (c) and determined the largest possible value of $k$ and $s$ that lets one still implement such FE formulation. The software implementation of these algorithms is available at FEMPAR [20], an open source OO Fortran200X scientific software package for the HPC simulation of complex multiphysics problems governed by PDEs at large scales.

Besides, we have carried out a comprehensive strong scaling study of FEMPAR on a petascale supercomputer when applied to both Poisson and Maxwell PDE problems, for problems in which the usage of AMR is highly beneficial for computational efficiency. The study reveals remarkable scalability of the framework up to several tens of thousands of CPU cores in the solution of problems with several hundreds of millions of degrees of freedom. Besides, we have also compared performance and strong scalability of FEMPAR against the deal.II FE library. This comparison reveals that the algorithms proposed here are at least as fast deal.II, and, in many cases, *2-3 times faster*.

The usage of parallel scalable AMR is still rather limited by most scientific computing practitioners and researchers, mostly because of the underlying complexity behind developing such tool. We expect this paper to be successful in convincing the reader that the design and development of the algorithms and data structures behind this tool is affordable and of reasonable complexity. With this in mind, while still using a high-level approach when presenting the algorithms, we introduced sufficient discussion and associated details to be helpful in this regard. Besides, wherever applicable, and in contrast to the other related work in the literature, the algorithms are supplemented with mathematical propositions and proofs that support their correctness.

APPENDIX A. PROOFS OF PROPOSITIONS

*Proof of Prop. 2.7.* By Def. 2.5, for any $K' \in \mathcal{T}_{K,F}$, it holds $F \in [\mathcal{F}_{K'}]$ and $\overline{F} \subset \overline{K} \cap \overline{K'}$. If $\ell(K) > \ell(K')$ were true, then there would be an ancestor $K''$ of $K$ at the same refinement level of $K'$. By Ass. 2.6, the mesh composed of all leaves of the refinement tree $\mathcal{S}_{\ell(K')}$ is a conforming mesh. Thus, by Def. 2.1, there exists an $n$-face $G \in [\mathcal{F}_{K'}] \cap [\mathcal{F}_{K''}]$ such that $\overline{G} = \overline{K'} \cap \overline{K''}$. Since $\overline{F} \subset \overline{K} \cap \overline{K'}$, $\overline{G} = \overline{K'} \cap \overline{K''}$, $K \subset K''$, and the refinement rule subdivides all cell $n$-faces with $n > 0$ (see Ass 2.2), then $F \subsetneq G$. As a result, $K'$ has two *different* VEFs $F, G \in [\mathcal{F}_{K'}]$ such that $F \cap G \neq \emptyset$. This is a contradiction, since the VEFs of a cell are disjoint by definition. We proceed analogously for $\ell(K) < \ell(K')$. It proves the result for $\mathcal{T}_{K,F}$. The result for $\mathcal{T}_{K,F}^-$ and $\mathcal{T}_{K,F}^+$ can be proved in a similar way. □

**Proposition A.1.** *(Intermediate result) For any forest-of-trees $\mathcal{T}$, $F \in \mathcal{F}$, $K \in \mathcal{T}_F$, and $K' \in \tilde{\mathcal{T}}_F$, it holds $\ell(K) > \ell(K')$.*

*Proof of Prop. A.1.* Cell levels must be different due to the conformity in Ass. 2.6. Let us assume that $\ell(K) < \ell(K')$. We can consider the recursive refinement of $K$ till reaching $\ell(K')$. In this process, by the definition of the refinement rule, any $n$-face of $K$ with $n > 0$ is being partitioned into a set of VEFs at level $\ell(K')$. By the conformity of the uniformly refined tree at $\ell(K')$, any VEF in $\mathcal{F}_{K'}$ in touch with $\overline{K}$ is a strict subset of a VEF in $\mathcal{F}_K$ or a vertex in $\mathcal{F}_K$. Thus, the situation in the statement is not possible and $\ell(K) > \ell(K')$ must hold. □

*Proof of Prop. 3.2.* Let us represent the right-hand side of (1) (resp., (2)) with $\mathcal{S}_F$ (resp., $\tilde{\mathcal{S}}_F$). We want to prove that $\mathcal{T}_F = \mathcal{S}_F$ and $\tilde{\mathcal{T}}_F = \tilde{\mathcal{S}}_F$. First, we note that $\mathcal{T}_F \cup \tilde{\mathcal{T}}_F$ is the set of all cells $K' \in \mathcal{T}$ such that $F \subset \overline{K'}$. On the other hand, $\mathcal{S}_F \cup \tilde{\mathcal{S}}_F = \{K\} \cup \mathcal{S}_{K,F} \cup \mathcal{S}_{K,F}^+ \cup \mathcal{S}_{K,F}^-$ spans the same set of cells, which can be proved using Def. 2.5. Thus, $\mathcal{T}_F \cup \tilde{\mathcal{T}}_F = \mathcal{S}_F \cup \tilde{\mathcal{S}}_F$. On the other hand, one can check that $\mathcal{S}_F \subset \mathcal{T}_F$, since: (a) $K \in \mathcal{T}_F$ by hypothesis, (b) $\mathcal{S}_{K,F} \subset \mathcal{T}_F$ by Def. 2.5, and (c) $\mathcal{S}_{K,F}^+$ and the fourth set in the definition of $\mathcal{S}_F$ are subsets of $\mathcal{T}_F$ by definition. One can also readily check that $\tilde{\mathcal{S}}_F \cap \mathcal{T}_F = \emptyset$ by its definition. Combining these results, we prove that $\mathcal{T}_F = \mathcal{S}_F$ and $\tilde{\mathcal{T}}_F = \tilde{\mathcal{S}}_F$. □

*Proof of Prop. 3.3.* Let us prove that for $n$-faces $F$ with $n > 0$, it holds: (1) $\mathcal{S}_{K,F}^+ = \emptyset$ and (2) $\{K' \in \mathcal{S}_{K,F}^- : F \in [\mathcal{F}_{K'}]\} = \emptyset$ and $\{K' \in \mathcal{S}_{K,F}^- : F \notin [\mathcal{F}_{K'}]\} = \mathcal{S}_{K,F}^-$. We prove (1) using the same argument as in Prop. A.1; a cell $K' \in \mathcal{S}_{K,F}^+$ belongs to $\mathcal{T}_{K,G}^+$ for $G \in [\mathcal{F}_K]$. It holds $\ell(K') > \ell(K)$ by Prop. 2.7, since $G$ is an $n$-face with $n > 0$. On the other hand, $F \notin [\mathcal{F}_K] \cap [\mathcal{F}_{K'}]$ by using the fact that $n$-faces with $n > 0$ are partitioned with refinement. Thus, $\mathcal{S}_{K,F}^+$ is empty. The same arguments can be readily used to prove (2). Combining these results and Prop. 3.2, we end the proof. □

**Proposition A.2.** *(Intermediate result) Let us consider a 2:1 $k$-balanced mesh $\mathcal{T}$ and an $n$-face $F \in \mathcal{F}_H$ with $n \geq k$. For any $K \in \mathcal{T}$ with a local VEF $f \in \mathcal{F}_K$ satisfying $[f] = F$, there exists $K' \in \mathcal{T}$ and $g \in \mathcal{F}_{K'}$ such that $[g] = [O_f]$. Thus, one can define the owner VEF of $F$ as $O_F \doteq [g]$ and $F \subsetneq O_F$. If $F$ is a vertex, it also holds for a 2:1 1-balanced mesh.*

*Proof of Prop. A.2.* Due to Prop. 2.7 and the 2:1 $k$-balance, since the VEF $F$ is hanging, it is in touch with a cell $K'$ at level $\ell(K) - 1$. By definition, $O_f \in \mathcal{F}_{K_p}$ where $K_p$ is the parent cell of $K$. By conformity of the uniformly refined tree at level $\ell(K) - 1$, there exists a $g \in \mathcal{F}_{K'}$ such that $[O_f] = [g]$. Thus, one can define $O_F$ as $[g] \in \mathcal{F}$.

If $F \in \mathcal{F}_H$ is a vertex, there is a $K' \in \mathcal{T}$ such that $F \subset \overline{K'}$ but $F \notin [\mathcal{F}_{K'}]$. For any $K \in \mathcal{T}$, $f \in \mathcal{F}_K$ such that $[f] = F$, we consider the ancestor $K_p$ of $K$ at level $\ell(K')$. By conformity of the uniformly refined tree, we can pick $g \in \mathcal{F}_{K_p}$ and $g' \in \mathcal{F}_{K'}$ such that $\overline{K_p} \cap \overline{K'} = \overline{g} = \overline{g'}$. Clearly, $f \subsetneq g'$ (otherwise $f \in \mathcal{F}_{K'}$) and thus, $g$ and $g'$ are $n$-faces with $n > 0$. By the 2:1 1-balance, $\ell(K') = \ell(K_p) = \ell(K) - 1$. As a result, $O_f \in \mathcal{F}_{K_p}$ (i.e., $K_p$ is the parent cell), $O_f$ is a VEF of $g$ (by the polytope definition), and there exists an $h \in \mathcal{F}_{K'}$ such that $[h] = [O_f]$. Therefore, $[O_f] \in \mathcal{F}$ and we can define $O_F \doteq [O_f]$. □

*Proof of Prop. 3.5.* It is a direct consequence of the construction of $O_F$ in Prop. A.2. □

*Proof of Prop. 3.6.* If $F \in \mathcal{F}_H$, there exists a $K \in \mathcal{T}_F$ and a $K' \in \tilde{\mathcal{T}}_F$. If $O_F \in \mathcal{F}_H$, it would be in touch with a cell in level $\ell(K') - 1$ due to Prop. A.1. It contradicts the 2:1 $k$-balanced assumption. Thus, $O_F \in \mathcal{F}_R$. Finally, an $n$-face $E$ of $O_F$ belongs to $\mathcal{F}_{K'}$; the $n$-faces of an $n$-face of $K$ are $n$-faces of $K$ by construction of a polytope. Thus, it is easy to check that $E$ is in touch with $K'$ and a sibling cell of $K$. As above, if $E$ would be hanging, it would violate the 2:1 $k$-balance. □

*Proof of Prop. 3.7.* To prove the first result, we must show that $O_V \in \mathcal{F}_R$ for any vertex $V \in \mathcal{F}_H$. The result for $O_V$ being an n-face with $n > 0$ has been proved in Prop. 3.6. If $O_V$ would be a vertex, $V = O_V$ and thus $O_V \in \mathcal{F}_H$. Thus, $V$ would belong to a cell $K$ and its parent cell in $\ell(K) - 1$ would be in touch with a cell $K'$ in $\ell(K) - 2$ or lower in $\tilde{\mathcal{T}}_V$. Thus, there would be an $n$-face $G \in [\mathcal{F}_{K'}]$ such that $V \subsetneq G$, thus $n > 0$. Thus, $G$ would violate the 2:1 1-balance property. It proves the first result and also shows that hanging vertices are always 2:1 balanced. Using this fact, we can prove the second result as in Prop. 3.6. □

*Proof of Prop. 3.11.* Let us first consider the case $F \in \mathcal{F}_R^p \cap \mathcal{F}_L^p$. As $F \in \mathcal{F}_R^p$, then $\tilde{\mathcal{T}}_F^p = \emptyset$. Thus, $F \in \mathcal{F}_L^p$ because of (a), (c), or (d) in Def. 3.9 (i.e., it cannot be in $\mathcal{F}_L^p$ due to (b)). Let us assume that $F \in \mathcal{F}_L^p$ because of (a) in Def. 3.9, i.e., at least one of the cells in $\mathcal{T}_F^p$ is local. For the first proposition statement, as $F$ fulfills (a), then $\mathcal{T}_F = \mathcal{T}_F^p$ and $\tilde{\mathcal{T}}_F = \tilde{\mathcal{T}}_F^p = \emptyset$ for those $n$-faces $F$ for which $n \geq k$ (see discussion above where the $\mathcal{T}_F^p$ and $\tilde{\mathcal{T}}_F^p$ sets are defined). For the second proposition statement, i.e., 0-faces $F$ and $k = 1$, we have that $\tilde{\mathcal{T}}_F = \tilde{\mathcal{T}}_F^p = \emptyset$ due to Prop. 3.2. Thus, $F \in \mathcal{F}_R$ in both cases. If $F \in \mathcal{F}_R^p$ is in $\mathcal{F}_L^p$ because it holds (c) or (d) in Def. 3.9, $F \in \mathcal{F}_R$ due to Prop. 3.6 in the case of the first proposition statement, and due to Prop. 3.7 in the case of the second. On the other hand, if $F \in \mathcal{F}_R \cap \mathcal{F}_L^p$ it is clearly in $\mathcal{F}_R^p \cap \mathcal{F}_L^p$, thus $\mathcal{F}_R^p \cap \mathcal{F}_L^p = \mathcal{F}_R \cap \mathcal{F}_L^p$. Using the fact that both $\{\mathcal{F}_R^p \cap \mathcal{F}_L^p, \mathcal{F}_H^p \cap \mathcal{F}_L^p\}$ and $\{\mathcal{F}_R \cap \mathcal{F}_L^p, \mathcal{F}_H \cap \mathcal{F}_L^p\}$ are partitions of $\mathcal{F}_L^p$, we readily have that $\mathcal{F}_H^p \cap \mathcal{F}_L^p = \mathcal{F}_H \cap \mathcal{F}_L^p$. □

*Proof of Prop. 4.1.* The fact that the constraints are direct is an immediate consequence of how the constraints are defined [12], Prop. 3.6, and (for $D(\mathcal{V}_h) = 0$) Prop. 3.7. On the other hand, for DOFs owned by VEFs $F \in \mathcal{F}_H \cap \mathcal{F}_L^p$, then $O_F$, and any of its boundary VEFs are in $\mathcal{F}_L^p$, due to Def. 3.9, i.e., such DOFs are only constrained by DOFs owned by $\mathcal{F}_R \cap \mathcal{F}_L^p$. Since $k$-ghost cells are locally accessible, then $O_F$, and any of its boundary VEFs are locally accessible, and thus the constraints on DOFs owned by such VEFs $F$ can be computed locally. □

*Proof of Prop. 4.2.* If $D(\mathcal{V}_h) \geq s$, then for $g \in \mathcal{N}_R^p \cap \mathcal{N}_\Gamma^p$ such that $g \in \mathcal{N}_F^p$ with $\mathcal{T}_F^p \cap \mathcal{T}_L^p \neq \emptyset$, then $\mathcal{T}_F^p = \mathcal{T}_F$ (see Sect. 3.2). Thus, all processors in $\mathcal{S}^{\text{proc}}(g)$ such that $g$ is surrounded by at least one local cell can compute $\mathcal{O}^{\text{proc}}(g)$ locally. The last part of the proposition is straightforward. □

*Proof of Prop. 4.3.* Given a processor $q \in \mathcal{S}^{\text{proc}}(g)$, by definition, the VEF $F$ that owns $g$ must be in $\mathcal{F}_L^q$ because of (a), (c), or (d) in Def. 3.9. Using the definition of $k$-ghost cells in Def. 3.8, when $F$ satisfies (a) or (c) at processor $q$, the fact that $q$ is in $\mathcal{S}^{\text{proc}}(g)$ can be locally determined at $\mathcal{O}^{\text{proc}}(g)$. When (d) holds for some VEF $J$ such that $F \subset \bar{J}$ (where $J$ is regular by Prop. 3.7), it is easy to check that $\mathcal{O}^{\text{proc}}(J)$ (which cannot be $q$, as $J$ is only surrounded by ghost cells in $q$) is neighbor of $\mathcal{O}^{\text{proc}}(g)$ and can determine that $q \in \mathcal{S}^{\text{proc}}(g)$. Thus, $q$ can be fetched at $\mathcal{O}^{\text{proc}}(g)$ with a single nearest neighbor communication. □